\documentclass[11pt,a4paper]{article}
\usepackage[hmargin=1.1in,vmargin={1.1in,1.1in}]{geometry}
\PassOptionsToPackage{hyphens}{url}\usepackage[colorlinks]{hyperref}
\usepackage{amsmath}
\usepackage{multirow}
\usepackage{mathtools}
\usepackage{amssymb}
\usepackage{dsfont}
\usepackage{bm}
\usepackage{url}
\usepackage{float}
\usepackage{tabularx}

\usepackage{natbib}

\usepackage[font=footnotesize]{caption,subcaption} 
\usepackage{graphicx}
\usepackage[affil-it]{authblk}
\usepackage{array} 
\usepackage{verbatim}
\usepackage{placeins} 
\usepackage[dvipsnames,table,xcdraw,svgnames]{xcolor}

\newcolumntype{L}[1]{>{\let\newline\\\arraybackslash\hspace{0pt}}m{#1}}

\newtheorem{algo}{Algorithm}
\newcommand{\matr}[1]{\mathbf{#1}} 

\AtBeginDocument{%
 \hypersetup{
  citecolor=Bittersweet,
    linkcolor=Blue,
  urlcolor=Blue}
}

\title{ 
Simultaneous supply and demand constraints in input-output networks: The case of Covid-19 in Germany, Italy, and Spain
 }

\author{Anton Pichler$^{1,2,3}$ and
	J. Doyne Farmer$^{1,2,4}$ \\

\vspace{0.25cm}

\footnotesize{$^{1}$ Institute for New Economic Thinking at the Oxford Martin School, University of Oxford, UK} \\
\footnotesize{$^{2}$ Mathematical Institute, University of Oxford, UK}\\
\footnotesize{$^3$ Complexity Science Hub Vienna, Austria} \\
\footnotesize{$^{4}$ Santa Fe Institute, US}\\
}

\date{\today}

\begin{document}

\maketitle

\begin{abstract}
	\noindent
	Natural and anthropogenic disasters frequently affect both the supply and demand side of an economy. A striking recent example is the Covid-19 pandemic which has created severe disruptions to economic output in most countries. These direct shocks to supply and demand will propagate downstream and upstream through production networks. Given the exogenous shocks, we derive a lower bound on total shock propagation. We find that even in this best case scenario network effects substantially amplify the initial shocks. To obtain more realistic model predictions, we study the propagation of shocks bottom-up by imposing different rationing rules on industries if they are not able to satisfy incoming demand. Our results show that economic impacts depend strongly on the emergence of input bottlenecks, making the rationing assumption a key variable in predicting adverse economic impacts. We further establish that the magnitude of initial shocks and network density heavily influence model predictions. \\[.3cm]
\end{abstract}

\vfill

\footnotesize{ 
\noindent
\emph{Acknowledgments:}
We thank F. Lafond for helpful feedback.
This work was supported by Baillie Gifford, Partners for a New Economy, the UK's Economic and Social Research Council (ESRC)  via the Rebuilding Macroeconomics Network (Grant Ref: ES/R00787X/1), the Oxford Martin Programme on the Post-Carbon Transition, and the Institute for New Economic Thinking at the Oxford Martin School.
This research is based upon work supported in part by the Office of the Director of National Intelligence (ODNI), Intelligence Advanced Research Projects Activity (IARPA), via contract no. 2019-1902010003. The views and conclusions contained herein are those of the authors and should not be interpreted as necessarily representing the official policies, either expressed or implied, of ODNI, IARPA, or the U.S. Government. The U.S. Government is authorized to reproduce and distribute reprints for governmental purposes notwithstanding any copyright annotation therein.
\\
\noindent
\emph{Corresponding author:}
Anton Pichler (anton.pichler@maths.ox.ac.uk)
}
\normalsize

\newpage
\section{Introduction} \label{sec:intro}

An immanent feature of many natural and anthropogenic disasters is that they affect both the supply and the demand side of the economy. In this paper we study the Covid-19 pandemic as an exemplary case of simultaneous supply and demand shocks.
Supply shocks from the pandemic arise from different sources. While deaths and sickness of employees can limit productive capacity, these effects are minor compared to nation-wide lockdown measures imposed by governments to curb the spreading of the virus. During lockdown workers employed in non-essential industries who cannot work remotely are unable to perform their jobs \citep{del2020supply, dingel2020, koren2020business}.

The pandemic also affects demand in heterogeneous ways \citep{CBO2006, del2020supply, chetty2020did, carvalho2020tracking, chen2020impact}.
While we expect demand shocks to be comparatively small for some industries (e.g. manufacturing), other industries are strongly affected by demand shocks. An illustrative example is the transportation industry which is considered as essential in many countries and thus would not experience adverse supply side effects. But the transportation industry faces large demand shocks, since consumers reduce demand for air travel and public transport to avoid infectious exposure.

Since economic agents are embedded in production networks, we expect that the overall economic impact is larger than the initial shocks to supply and demand suggest. Demand shocks will reduce the sales of firms and, by backward propagation, also diminish the sales of their suppliers.
Supply shocks, on the other hand, will spread downstream and upstream. Downstream effects materialize if the limited productive capacity of suppliers creates input bottlenecks for customers. Due to lower productive capacities, firms will also require less inputs for production, thus adversely affecting upstream suppliers of these inputs.

Input-output (IO) models are frequently used to model higher-order economic impacts arising from such exogenous shocks. Most of these models account for either supply or demand shocks but do not incorporate them concurrently \citep{galbusera2018input}. In this paper we revisit existing IO modeling techniques and introduce a novel dynamic approach, to account for both types of shocks.
Our analysis is based on industry-level data but could easily be extended to firm-level analysis that has become the focus of recent studies \citep{diem2021quantifying, borsos2020shock,mandel2020disequilibrium,inoue2019firm}. 
In particular, our results relate directly to the effects of data aggregation on impact estimates.

Several macroeconomic models that account for the production network structure have been suggested to study the economic impacts of the Covid-19 pandemic. \cite{inoue2020propagation} use an agent-based model calibrated to more than a million Japanese firms to model nation-wide economic effects of a lockdown in Tokyo. Using the World Input-Output Database (WIOD), \cite{mandel2020economic} study the effects of national lockdowns on global GDP in a non-equilibrium framework. \cite{fadinger2020effects, baqaee2020nonlinear, bonadio2020global} and \cite{barrot2020sectoral} use general equilibrium models to quantify economic impacts of social distancing. All these studies focus on the supply-side shocks of the pandemic and without further considering changes in final consumption. Exceptions are \cite{pichler2020production} and \cite{guan2020global} who incorporate both supply and demand shocks in their production network models.

In contrast to these studies, we consider simpler modeling techniques derived from traditional IO analysis which we extend to account for simultaneous supply and demand shocks. We follow this approach to better isolate underlying mechanisms of shock propagation in production networks that can be hard to disentangle in more sophisticated macroeconomic models.
When applying pandemic shocks to data from Germany, Italy and Spain, we find that existing IO modeling techniques yield infeasible solutions of economic impacts. To understand the nature of this problem we introduce a simple linear programming method that allows us the determine feasible market allocations, representing a lower bound of minimal shock propagation. To find more realistic impact estimates, we further introduce a new bottom-up approach based on rationing outputs. This setup allows us to dynamically model the propagation of shocks in production networks and uncover several theoretical implications, such as the effect of alternative behavioral assumptions on economic impacts.

We intentionally keep our modeling approach simple.
As mentioned above, even more complicated models currently applied to assess the economic effects of the pandemic do not incorporate supply and demand shocks simultaneously.
The focus of this paper is on the theoretical implications of simultaneous supply and demand constraints for IO-based impact assessment methods.
For more realistic empirical impact assessment of natural disasters, we stress that further aspects that we do not consider here could play an important role, e.g. inventory dynamics \citep{bak1993aggregate, pichler2021lockdown}, price and substitution effects, adaptive behavior and governmental intervention (see \cite{oosterhaven2017limited} for a recent discussion).

Our paper contributes to the field of IO models in disaster impact assessment, for which a recent review can be found in \cite{galbusera2018input}. For recent demand-side and supply-side focused contributions see \cite{klimek2019quantifying} and \cite{avelino2019comparing}, respectively. Similar to our derivation of a lower bound of shock propagation, \cite{koks2016multiregional} and \cite{oosterhaven2016new} rely on optimization techniques to account for supply constraints in demand-driven models. Constrained optimization techniques are frequently used in extended IO models (for example \cite{duchin2005world} and \cite{duchin2012rectangular}).
The modeling of alternative rationing algorithms is related to the studies of \cite{steenge2007thinking, li2013modeling} and \cite{koks2015integrated} which allow for imbalanced IO tables immediately after the disaster strikes and evaluate different recovery paths.

Our findings show that shock amplification is much larger in the bottom-up approaches than what best case scenarios would suggest.  Micro-level coordination failures can severely exacerbate adverse economic shocks. 
Rationing assumptions play a key role in impact estimates and alternative behavioral rules can lead to very different aggregate outcomes. These effects strongly interact with the overall shock magnitude as well as the level of connectedness in the production network. 
While we find that economic impacts increase with higher levels of network density in general, the extent to which this is true strongly depends on the underlying rationing mechanisms.
Our results also imply that estimates of economic impact can be highly sensitive with respect to data quality and aggregation.

This paper is organized as follows. We first discuss first-order pandemic shocks to supply and demand, as well as the datasets used (Section \ref{sec:FOshocks}). In Section \ref{sec:meem} we introduce the basic IO framework and 
discuss how existing approaches have difficulties in incorporating simultaneous supply and demand constraints.
Section \ref{sec:shockprop} discusses the main results of this paper.
We propose an optimization method in Section \ref{sec:optim} and introduce alternative rationing algorithms in Section \ref{sec:ration} to model shock propagation in supply and demand constrained production networks. We show empirical results and discuss further theoretical implications in Section \ref{sec:results} before concluding in Section \ref{sec:discuss}.

\section{Pandemic shocks to supply and demand} \label{sec:FOshocks}

Our analysis is based on the most recent year (2014) of the World-Input-Output-Database (WIOD) \citep{timmer2015illustrated}.\footnote{
	All code and data to reproduce this paper are made accessible online: \url{https://www.doi.org/10.5281/zenodo.4326815}.
}
We use estimates of supply and demand shocks from the literature to determine maximum final consumption and production values for 54 industries in three major European economies, i.e. Germany, Italy and Spain.\footnote{
	We removed industries \emph{T} (Household activities) and \emph{U} (Extraterritorial activities) from our analysis since they are not connected to other industries in the data and thus don't play any role in the propagation of shocks in the production network. 
}
 We focus on these economies because existing research provides us with lists of essential industries and thus allows us to make reasonable estimates of supply shocks. For this reason, we only allow for domestic supply shocks in our analysis, despite acknowledging the importance of international supply chain disruptions.

We follow the approach of \cite{del2020supply} to compute supply shocks for every industry during a lockdown. A supply shock is caused by the removal of labor in non-essential industries due to social distancing measures.  In contrast, an industry which is defined as essential will not be affected. Even if employed in non-essential industries, workers who can accomplish their work from home will do so and are assumed to keep pre-lockdown productivity levels. The Remote Labor Index $\text{RLI}_i$ indicates the share of an industry's labor force that can work from home.
The supply shock to industry $i$ is then computed as $\epsilon_i^S = (1-\text{RLI}_i)(1-\text{e}_i)$, where $\text{e}_i$ is the share of an industry defined as essential. 
Thus, the supply shock of an industry during lockdown is the share of labor in non-essential industries that cannot work from home. 
We assume that the supply shock to an industry determines the maximum production of an industry, i.e.
\begin{equation}
	x_{i}^\text{max} = (1-\epsilon_i^S) x_{i,0},
\end{equation}
where $x_{i,0}$ denotes the production in industry $i$ before lockdown. Following this approach, every industry faces binding supply side constraints, limiting its production to $x_i \in [0,x_{i}^\text{max}]$.
The Remote Labor Index is taken from \cite{del2020supply} and industry-specific essential ratings for Germany, Italy and Spain are obtained from \cite{fana2020covid}.

To determine first-order demand shocks to every industry, $\epsilon_i^D$, we use estimates of a prospective Congressional Budget Office (CBO) study aiming at quantifying the demand-side impact of a pandemic \citep{CBO2006}. Demand and supply shock data have been mapped to WIOD industry categories by \cite{pichler2021lockdown} and are presented in detail in Appendix \ref{apx:shocks}.
As for supply shocks, we assume that demand shocks determine maximum final consumption values according to
\begin{equation}
f_{i}^\text{max} = (1-\epsilon_i^D) f_{i,0},
\end{equation}
where $f_{i,0}$ represents pre-lockdown final consumption. Since consumption cannot be negative, we have $f_i \in [0,f_{i}^\text{max}]$.\footnote{
In principle, $f_i$ could be negative if there is extremely large inventory depletion. This is not the case in our data. In national accounts inventory adjustment merely represents a variable to rebalance row and column sums of IO tables. We therefore do not consider the possibility of negative final demand.
}
Note that WIOD distinguishes different final consumption categories.\footnote{
Given the multi-regional nature of WIOD, imports from and exports to other industries are accounted for in the intermediate consumption matrix. Since we treat international trade as exogenous, we aggregate all intermediate exports to a single final consumption category, as is usually the case with national IO tables. Similarly, we treat intermediate imports as an exogenous input that does not inhibit production, or equivalently, that is always available if demanded. 
} 
We apply the CBO estimates to final consumption of households and non-profit organizations and assume a 10\% shock to investments and exports as in \cite{pichler2021lockdown}. 
Due to our methodological focus we do not further distinguish between different final consumption categories but for simplicity only consider total final consumption values for every industry. It is important to point out that this is a major simplification and we expect the various types of final consumption to be highly relevant in empirical assessments.  
Table \ref{t:aggregate} shows country aggregates of essentialness scores, Remote Labor Index, supply and demand shocks.

\begin{table}[ht]
	\centering
	\begin{tabular}{|r|rrrrrr|}
		\hline
		& $x$ & $f$ & $\epsilon^S$ & $\epsilon^D$ & RLI & $e$ \\ 
		\hline
		Germany & 7,066.74 & 4,447.11 & 0.31 & 0.09 & 0.42 & 0.49 \\ 
		Italy & 4,075.40 & 2,343.12 & 0.27 & 0.11 & 0.41 & 0.55 \\ 
		Spain & 2,567.91 & 1,552.88 & 0.33 & 0.13 & 0.40 & 0.44 \\ 
		\hline
	\end{tabular}
\caption{ \textbf{Country aggregates of gross output, final consumption and supply and demand shock inputs.} Columns $x=\sum_i x_{i,0}$ and $f=\sum_i f_{i,0}$ are country gross output and final consumption in billion USD based on 2014 values. $\epsilon^S = 1 - \sum_i x_i^\text{max}/x$ and $\epsilon^D = 1 - \sum_i f_i^\text{max}/f$ are total supply and demand shocks, respectively. $\text{RLI}=\sum_i \text{RLI}_i x_{i,0} / x$ and $e= \sum_i e_i x_{i,0} / x$ denote the country-wide Remote Labor Index and essentialness score on which supply shocks are based. 
In all three countries total direct supply shocks are larger than demand shock.	
}
\label{t:aggregate}
\end{table}

\section{IO framework} \label{sec:meem}
We first introduce basic national accounting identities that build the basis for most IO models. Let us consider an economy consisting of $n$ industries, each producing a unique, industry-specific good. Inter-industry purchases and sales are encoded in the intermediate consumption matrix $\matr Z$ where an element $z_{ij}$ denotes the total monetary value of goods produced by industry $i$ that are consumed by industry $j$. For the $n$-dimensional vectors of gross output, total final consumption and value added, we write $\bm x$, $\bm f$ and $\bm v$, respectively. In this economy the following identities hold
\begin{align} \label{eq:io_account}
\bm x = \matr Z \mathbf{i} + \bm f = \matr Z^\top \mathbf{i} + \bm v, 
\end{align}
where $\mathbf{i}$ is a $n$-dimensional column vector of ones.

A core assumption in a Leontief-inspired modeling framework is that industries produce based on fixed production recipes, allowing us to rewrite the first identity of Eq.~\eqref{eq:io_account} as
\begin{equation} \label{eq:leontief}
\bm x = \matr A \bm x + \bm f = \matr L \bm f,
\end{equation}
where $\matr A$ is the technical coefficient matrix with elements $a_{ij} \equiv z_{ij}/x_j$ (the production recipe) and $\matr L$ the Leontief inverse $\matr L \equiv (\mathbf{I} - \matr A)^{-1}$. A conventional assumption is that gross output is determined endogenously, whereas final consumption is taken to be as exogenous. It then follows that value added is obtained as a residual variable.

Eq.~\eqref{eq:leontief} is at the core of many IO models that are frequently used for disaster impact analysis. This specification defines a demand-driven model.  However, as we have already stressed, many economic shocks actually act on the supply side of the economy. For example, natural catastrophes such as earthquakes or floods destroy physical capital, putting upper limits on an industry's production in the disaster aftermath.  
Eq.~\eqref{eq:leontief}, however, neglects supply capacity constraints and implies an infinite elasticity of supply with respect to demand. This is particularly problematic given the frequent short-term focus of IO studies.

We should point out that there are also supply-driven IO models building upon Ghosh's model \citep{ghosh1958input} that assumes exogenous primary factors and derives final consumption values endogenously. Ghosh's model does not comply with a Leontief production function and has been heavily criticized amongst other aspects for the assumption of perfect demand elasticities and perfect substitution of inputs \citep{oosterhaven1988plausibility, gruver1989plausibility, de2009ghosh}. 
The IO inoperability model \citep{haimes2001leontief,santos2004modeling} is another notable supply-side focused model. \cite{dietzenbacher2015reflections} show, however, that it closely corresponds to conventional IO modeling approaches.

Neither the demand- nor the supply-driven specification of IO models are able to incorporate supply and demand constraints at the same time. A potential remedy is the mixed endogenous/exogenous model (MEEM), which has been applied in several studies including \cite{steinback2004using, kerschner2009erratum} and \cite{arto2015global}. The MEEM acknowledges that not only final demand is constrained but that for some industries supply constraints are more severe and therefore binding. 
A difficulty in empirical analysis is to define which industries are supply and which are demand constrained. In some cases there might be ``natural'' distinctions. For example, \cite{arto2015global} study global supply chain disruption effects of the 2011 T\={o}hoku earthquake by solving the MEEM where only the Japanese transport equipment industry is supply constrained, whereas for all other industries gross output is endogenous. In case of simultaneous (severe) supply and demand shocks the line of distinction will be more blurred.

Note that the MEEM incorporates both supply and demand shocks, but not \emph{simultaneously} for a single industry. Instead, an industry is either supply or demand constrained. As a consequence, solutions to the MEEM might not comply with exogenous industry-level constraints to supply and demand, i.e. they can be economically \emph{infeasible}. As is shown in detail in Appendix \ref{apx:meem}, this is exactly what happens if we apply the MEEM to the pandemic shocks discussed in the previous section. For all three countries, the MEEM yields final consumption values that are either negative or larger than the exogenous constraints.

\section{Propagation of simultaneous supply and demand shocks} \label{sec:shockprop}

The industry-specific effects to supply and demand during the Covid-19 pandemic and the difficulty of incorporating them in the existing models motivated us to explore possible extensions of the IO modeling framework.  Rather than using more complicated models that can deal with simultaneous supply and demand shocks, such as computable general equilibrium (CGE) models or those discussed in the introduction, we intentionally keep the modeling approach simple.   We remain close to the Leontief framework, the ``work horse'' of many more advanced IO-based models. This allows us to better isolate the basic mechanisms of shock propagation that are easily conflated with other effects in more sophisticated models.

To determine lower bounds of shock propagation with respect to output and consumption, we study the idealized case of a social planner who allocates goods to maximize either total gross output or total final consumption within the given economic constraints. This will give us a best case scenario of negative economic impacts, i.e. the minimal decrease in total output and final consumption necessary to arrive within the set of feasible solutions given the exogenous constraints to supply and demand.
Of course, alternative objectives could be optimized as well (e.g. employment).

While deriving a lower bound is valuable, it is unlikely to be realistic.  We therefore consider a second approach by comparing alternative rationing algorithms. If an industry is supply constrained, it will not be able to satisfy all of its demand and thus needs to make a decision which customer to serve and to what extent. We implement several decision rules and investigate how this choice influences the estimated economic impact. In contrast to the optimization method or the MEEM, which simply compute an equilibrium, this approach explicitly computes the transient dynamics that lead to a new equilibrium.

\subsection{Feasible market allocations} \label{sec:optim}

Given exogenous constraints to supply and demand, what is the feasible market allocation that maximizes final consumption and/or total output?  The solution needs to lie within exogenous bounds on supply and demand and also needs to satisfy the assumption of Leontief production, Eq.~\eqref{eq:leontief}.
We seek market allocations
$\{ \bm x^*, \bm f^* \}$ that (a) respect given production recipes $\bm x^* = \matr L \bm f^*$ and (b) satisfy basic output and demand constraints $\bm x^* \in [ \bm 0, \bm x^\text{max}]$ and $ \bm f^* \in [ \bm 0, \bm f^\text{max}]$.

We follow a mathematical optimization procedure to map out the solution space of feasible market allocations.
As a first case, we determine the market allocation that maximizes gross output under the assumptions specified. Large levels of gross output indicate high economic activity, which in turn entail high levels of primary factors such as labor compensation. As a second case we look at market allocations that maximize final consumption given current production capacities. Due to the linearity of the Leontief framework, the problems boil down to linear programming exercises.

\paragraph{Maximizing gross output:}

\begin{align}   \label{eq:max_GO}
& \underset{ \bm f \in [\bm 0, \bm f^\text{max}] }{ \max }   \qquad \mathbf{i}^\top (\mathbf{I} - \matr A )^{-1} \bm f, \\
& \text{subject to}  \qquad (\mathbf{I} - \matr A )^{-1} \bm f \in [\bm 0, \bm x^\text{max}]. \nonumber
\end{align}

\paragraph{Maximizing final consumption:}

\begin{align}   \label{eq:max_FD}
& \underset{ \bm x \in [\bm 0,\bm x^\text{max}] }{ \max }   \qquad \mathbf{i}^\top (\mathbf{I} - \matr A ) \bm x, \\
& \text{subject to}  \qquad (\mathbf{I}- \matr A) \bm x \in [\bm 0, \bm f^\text{max}]. \nonumber
\end{align}

To maximize gross output of the economy, $\sum_i x_i^*$, requires us to the find the vector of final consumption $\bm f^*$. The constraint $\matr L \bm f \in [\bm 0, \bm x^\text{max}]$ ensures that industry output levels lie within the respective production capacities.
The problem is similar when maximizing final consumption where a vector of output levels $x^*$ is chosen to maximize final consumption, $\sum_i f_i^*$. The auxiliary constraint enforces that final consumption levels do not exceed given demand.  
The optimization problem always admits a solution since the trivial allocation of a full collapse $\{ \bm x^*, \bm f^* \} = \{ \bm 0, \bm 0\}$ always exists, although we expect positive values for realistic input data.

Note that the market allocations are ``optimal'' in a mathematical sense. That is, they represent maximum values of output and consumption given the exogenous constraints to supply and demand. Or stated differently, they determine the minimum level of shock propagation measured in aggregate final consumption and gross output. 
Thus, this optimization method is not intended to give realistic or necessarily desirable economic outcomes, but rather probes the system boundaries by mapping out the solution space. Any feasible solution, under the above-specified assumptions, has to lie within this space.

\subsection{Input bottlenecks and rationing variations} \label{sec:ration}

As our second method we implement different rationing schemes for output constrained industries. In contrast to the optimization methods, this represents a bottom-up approach for finding feasible market allocations. Industries place orders to their suppliers based on incoming demand. Since suppliers can be output constrained, they might not be able to satisfy demand fully. A supplier therefore needs to make a decision about how much of each customer's demand it serves. Intermediate consumers transform inputs to outputs based on fixed production recipes. Thus, if a customer receives less inputs than she asked for, she faces an input bottleneck further constraining her production. As a consequence, the customer reduces her demand for other inputs as they are not further needed under limited productive capacities. 
We iterate this procedure forward until the algorithm converges.

We run this algorithm with four alternative rationing rules: (a) strict proportional rationing, (b) proportional rationing to intermediate demand but priority of intermediate over final demand, (c) priority rationing serving largest customers first and (d) random rationing, where customers are served based on a random order. We then compare the results obtained from the four competing behavioral rules.

These rationing approaches are frequently applied in the literature, although there are differences in the exact specifications. In general it is hard to calibrate how firms distribute output in case of supply constraints to empirical data  and so the rationing choice is often ad-hoc. Here, we apply all four rules within a consistent dynamic framework to better understand how alternative behavioral assumptions affect impact estimates.
Strict proportional rationing is frequently assumed in the literature \citep{henriet2012firm, hallegatte2014modeling, guan2020global, mandel2020economic, pichler2020production} and also mixed proportional/priority rationing has been considered \citep{battiston2007credit, hallegatte2008adaptive, li2013modeling, diem2021quantifying}. The agent-based framework of \cite{inoue2019firm} and \cite{inoue2020propagation} assumes a variation of priority rationing while the random matching of suppliers and customers in the agent-based model proposed by \cite{poledna2018does} and \cite{poledna2019economic} most closely resembles a random rationing approach.

\paragraph{(a) Strict proportional rationing.}
If industries are unable to satisfy total incoming demand completely, they distribute output proportional to their customers' demand, where no distinction is made between intermediate and final customers.
More specifically, if an industry's output, $x_i$, is smaller than incoming demand, $d_i$, it will supply $z_{ij} = o_{ij} \frac{x_i}{d_i}$ to customer $j$, where $o_{ij}$ denotes the demand from customer $j$ to industry $i$. 
We implement the rationing algorithm in the following way:  First, industries determine their total demand as if there were no supply-side constraints, i.e. $\bm d = \matr L \bm f^\text{max}$. Industries then evaluate if they are able to satisfy demand given their constrained production capacities. If an industry $i$ can satisfy demand only partially, it will create a bottleneck of size $r_i = \frac{x^\text{max}_i}{d_i}$ to other industries due to proportional rationing. Since industries produce according to fixed Leontief input recipes, their largest input bottleneck, $s_i = \underset{j: a_{ji}>0}{\min} \{ r_j, 1 \}$ will be the binding constraint in production.
Thus, in case of input bottlenecks where $s_i<1$, production of $i$ reduces to $x_i =  \underset{ j: a_{ji}>0 }{ \min } \left \{ x_i^\text{max},  \frac{s_i a_{ji} d_i  }{a_{ji}} \right \} = \min  \{ x_i^\text{max}, s_i d_i \} < d_i$. This in turn reduces the amount of goods delivered to the final consumer $f_i = \min \{ x_i - \sum_j a_{ij} x_j, 0 \}$. 
The new final demand vector $\bm f$ now implies a new, lower level of aggregate demand, $\bm d = \matr L \bm f$, and we again let industries evaluate whether they can satisfy total demand within given production constraints. We iterate this procedure forward until all demand is met and no input bottleneck further constrains production.  
We can write the proportional rationing algorithm more compactly as follows:

\begin{algo}
Proportional rationing; industries are not prioritized over the final consumer. Take an initial demand vector $\bm f[0] = \bm f^\text{max}$ as given, implying an initial aggregated demand vector $\bm d[1] = \matr L \bm f[0]$. By looping over the index $t = \{1,2,...\}$, the following system is iterated forward:
\begin{align}
r_i[t] &= \frac{x_i^\text{max}}{d_i[t]},	\label{eq:prop_scon} \\
s_i[t] &= \underset{j: a_{ji}>0}{\min} \{ r_j[t], 1 \}, \label{eq:prop_bottle} \\
x_i[t] &= \min \{  x_i^\text{max}, s_i[t] d_i[t] \} , \label{eq:prop_production}\\
f_i[t] &= \max \left \{ x_i[t] - \sum_j a_{ij} x_j[t], 0 \right \}, \label{eq:prop_finald} \\
d_i[t+1] &=  \sum_j l_{ij} f_j[t]. \label{eq:prop_demand}
\end{align}
The algorithm converges to a new feasible economic allocation if $d_i[t+1] = d_i[t]$ for all $i$.
In this case output and final consumption levels are given as $x_i = d_i[t+1] = x_i[t+1]$
and $f_i = f_i[t+1]$, respectively.
\end{algo}

Eq.~\eqref{eq:prop_scon} indicates whether an industry is output constrained, where $r_i$ is the share of demand that can be met given existing productive capacities. 
If $r_i \ge 1$, industry $i$ is able to meet demand completely, whereas demand can only be partially satisfied if $r_i <1$. 
If any supplier of industry $i$ (which is the set  $\{j : a_{ji}>0 \}$) is sufficiently output constrained, industry $i$ faces an input bottleneck according to Eq.~\eqref{eq:prop_bottle}.
Due to perfect complementarity of inputs prescribed by the Leontief production function, industry $i$ can only produce a fraction of total demand as indicated in Eq.~\eqref{eq:prop_production}, reducing its delivery to final consumers as specified in Eq.~\eqref{eq:prop_finald}.   
The new total demand to an industry $i$ is then again derived through the weighted sum of final demand values where the weights are obtained from the Leontief inverse, Eq.~\eqref{eq:prop_demand}.

\paragraph{(b) Mixed proportional/priority rationing.}

It has been argued that firm-firm relationships are stronger than firm-household ties, so that intermediate demand should be prioritized over final demand \citep{hallegatte2008adaptive, inoue2019firm}.
While this assumption might make sense for households, this is not necessarily the case for other final demand categories. Final demand categories in national IO tables can include demand by governments and non-profit organizations, exports and investments and it is debatable whether intermediate demand should be prioritized over these categories. 

To quantify the effect of this assumption on the amplification of initial shocks, we implement a mixed proportional/priority rationing algorithm. Here, industries ration intermediate demand proportional analogously to the proportional rationing algorithm (a), but prioritize intermediate demand over final demand.
We outline this algorithm below.

\begin{algo}
	Proportional rationing among industries; industries are prioritized over final consumers.
	Take an initial demand vector $\bm f[0] = \bm f^\text{max}$ as given, implying an initial aggregated demand vector $\bm d[1] = \matr L \bm f[0]$. By looping over the index $t = \{1,2,...\}$, the following system is iterated forward:
	\begin{align}
	r_i[t] &= \frac{x_i^\text{max}}{\sum_j a_{ij} d_j[t]},	\label{eq:mix_scon} \\
	s_i[t] &= \underset{j: a_{ji}>0}{\min} \{ r_j[t], 1 \}, \label{eq:mix_bottle} \\
	x_i[t] &= \min \{ x_i^\text{max},  s_i[t] d_i[t]  \}, \label{eq:mix_production}\\
	f_i[t] &= \max \left \{ x_i[t] - \sum_j a_{ij} x_j[t], 0 \right \}, \label{eq:mix_finald} \\
	d_i[t+1] &=  \sum_j l_{ij} f_j[t]. \label{eq:mix_demand}
	\end{align}
	The algorithm converges to a new feasible economic allocation if $d_i[t+1] = d_i[t]$ for all $i$.
	In this case output and final consumption levels are given as $x_i = d_i[t+1] = x_i[t+1]$
	and $f_i = f_i[t+1]$, respectively.
\end{algo}

The algorithm is similar to the proportional rationing algorithm (a) but differs mainly in one aspect. Only intermediate demand affects the extent of an industry's output constraints, Eq.~\eqref{eq:mix_scon}. As a consequence, final demand does not play a role in the creation of input bottlenecks, Eq.~\eqref{eq:mix_bottle}.

\paragraph{ (c) Priority rationing (``largest first'').}

Since it is not obvious that industries should pass on their output proportionally in case they are not able to meet demand fully, we next consider a type of priority rationing.  Industries rank their customers based on demand magnitude and serve larger customers before smaller customers. 
In the proportional rationing setting an output constrained supplier affects all customers in the same way. Under a priority rationing scheme, however, supply shocks propagate downstream heterogeneously. For example, if a supplier cannot meet demand by only a small margin, most customers will not be affected by the priority rationing scheme. Only the smallest customers will face input bottlenecks, whereas every customer would experience the same small shock in the proportional rationing setup.

Intuitively, a priority rationing rule could make sense, as firms might have an interest in serving more important (large) customers fully, or at least as well as possible, before focusing on less important customers. It also seems closer to practice that firms process orders one-by-one instead of working through all orders simultaneously and leaving them incomplete to the same degree. While a priority rationing scheme could be plausible on the basis of firms or single transactions, it might be less so for more aggregate industry-level data. A link between two industries in IO tables corresponds to many firm/establishment level transactions and so it is not clear if large inter-industry links are due to a few big orders or many small orders. This becomes particularly evident when considering final consumers. Several industries face large demand from private consumers which effectively is the sum of many small orders (e.g. restaurants, grocery, theaters). Since we only consider an aggregate of final demand representing several distinct categories such as private consumers, government and investments, we exclude final demand from the priority rationing scheme. Thus, in the same manner as in the mixed prop./prior. rationing algorithm (b), we assume that intermediate demand is always prioritized over final demand.
By adopting this convention, we can formulate the priority rationing algorithm as follows.

\begin{algo} \label{algo3}
	Largest first rationing; industries are prioritized over the final consumer. 
	Take an initial demand vector $\bm f[0] = \bm f^\text{max}$ as given, implying an initial aggregate demand vector $\bm d[1] = \matr L \bm f[0]$. Every industry $i$ ranks each customers $j$ based on initial demand size: $h_{ij} = \{ k_{(1)},k_{(2)},...,k_{(j)} \, : \, a_{ik_{(1)}} d_{k_{(1)}}[1] \ge a_{ik_{(2)}} d_{k_{(2)}}[1]  \ge ... \ge a_{ik_{(j)}}  d_{k_{(j)}} [1] \}$.
	By looping over the index $t = \{1,2,...\}$, the following system is iterated forward:
	\begin{align}
	r_{ij}[t] &=  \frac{x_i^\text{max}}{ \sum_{n \in h_{ij} } a_{in_{(j)}} d_{n_{(j)}} [t] },	\label{eq:large_bottle} \\
	s_i[t] &= \underset{j: a_{ji}>0}{\min} \{ r_{ji} [t], 1 \}, \label{eq:large_scale} \\
	x_i[t] &= \min \{ x_i^\text{max},  s_i[t] d_i[t]  \}, \label{eq:large_production}\\
	f_i[t] &= \max \left \{ x_i[t] - \sum_j a_{ij} x_j[t], 0 \right \}, \label{eq:large_finald} \\
	d_i[t+1] &=  \sum_j l_{ij} f_j[t]. \label{eq:large_demand}
	\end{align}
	The algorithm converges to a new feasible economic allocation if $d_i[t+1] = d_i[t]$ for all $i$.
	In this case output and final consumption levels are given as $x_i = d_i[t+1] = x_i[t+1]$
	and $f_i = f_i[t+1]$, respectively.
\end{algo}

\paragraph{(d) Random rationing. }
As our final case, we again consider priority rationing. Rather than using a fixed ordering scheme based on demand magnitude, we use random priority. 
Industries rank their customers randomly and serve customers based on their position in the ranking. While largest-first rationing makes intuitive sense in some cases, it is unlikely to be a good approximation for all real-world settings. In practice, it is likely that other factors such as timing of orders matter. In that case industries could adopt a first-come-first-serve principle to process orders. Since IO data rarely comes with granular time information, we adopt a random ordering of incoming orders that mimics a first-come-first-serve principle under a uniform prior of which orders are coming in first. The algorithm is presented below.

\begin{algo}  \label{algo4}
	Random rationing; industries are prioritized over the final consumer. The algorithm is identical to Algorithm \ref{algo3}, except that the ranking of customer $j$ by industries $i$, $h_{ij} = \{ k_{(1)},k_{(2)},...,k_{(j)} \}$, is randomly drawn.
\end{algo}

While these algorithms are not guaranteed to converge to a steady-state equilibrium, we observe convergence in the vast majority of our simulation experiments. If the algorithm converges, the economic allocations obtained are automatically feasible.  This means that no industry has negative output or produces more than its productive capacities allow, final consumption is non-negative and below given exogenous maximum consumption levels, and there are no input bottlenecks left that further constrain production of downstream industries.  The Leontief equation $\bm x = \matr L \bm f$ holds, which implies that all sales add up to total output.

\subsection{Results} \label{sec:results}

We initialize these four rationing algorithms with the supply and demand shocks and IO data of Germany, Italy and Spain. We then compare the steady-state equilibrium economic outcomes predicted by the rationing algorithms with the optimization results and the direct shock computations (Appendix~\ref{apx:shocks}, Tables~\ref{t:ind_demand} and \ref{t:ind_supply}).

\begin{figure}[H]
	\centering
	\includegraphics[width=1\textwidth]{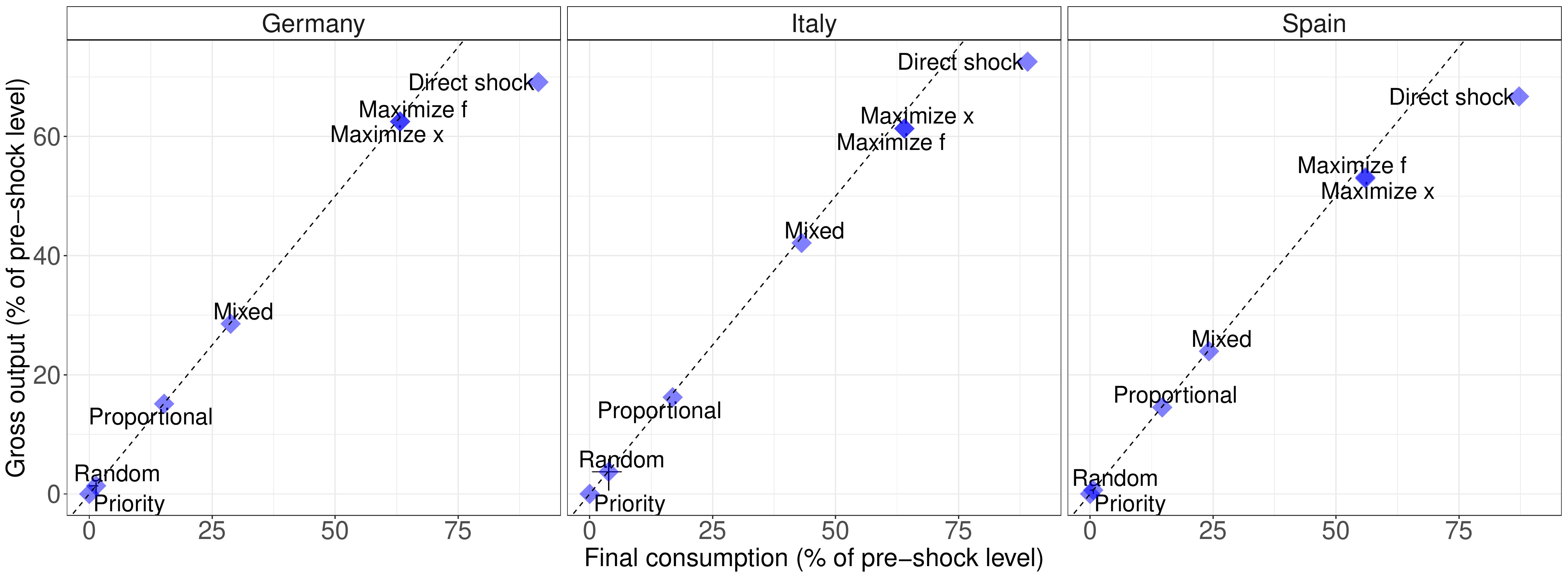}
	\caption{
		{\bf Comparison of different shock propagation mechanisms.} The y-axis denotes aggregate gross output levels normalized by pre-shock levels as predicted by the rationing and optimization methods. The x-axis shows the same for aggregate final consumption levels.  The \emph{Random rationing} diamond represents the average taken from 100 samples and the error bars indicate the interquartile range.
	}
	\label{fig:res_aggregate}
\end{figure}

Fig. \ref{fig:res_aggregate} summarizes the main result visually, where a diamond indicates the aggregate final consumption and gross output levels predicted by the different approaches. Note that a data point in the top-right corner indicates high gross output and final consumption values, i.e. limited negative impacts, whereas diamonds in the bottom-left corner represent extremely adversely impacted economies.
As already reported in Table \ref{t:aggregate}, for all three countries supply shocks are substantially larger than final consumption shocks, pushing the \emph{Direct shock} diamond substantially below the 45 degree line. Note that the direct shock market allocations are not feasible, as they ignore higher-order effects, such that a direct supply shock to an industry reduces the inputs of downstream industries and thus reduces their output too.

The best case feasible market allocations are given by the two optimization methods (maximize output vs. consumption), which yield exactly the same predictions on the aggregate and the industry level. This is true even though the two optimization methods are not equivalent.\footnote{
	When perturbing the economic systems we can find cases where the two optimization methods do not yield exactly the same results. Practically, we find that aggregate predictions are always similar for the two methods although industry-level results can sometimes differ significantly.
}
The two optimization methods have the highest output, but they still amplify the initial supply shocks to reduce the output from about $69\%$ to $63\%$ in Germany and from $67\%$ to $53\%$ in Spain.

All the other feasible market allocations, obtained from the rationing algorithms, lie substantially below the best case scenarios.
The wide range of predicted outcomes is the most striking aspect of Fig. \ref{fig:res_aggregate}.  Both the random rationing scheme and the priority rationing scheme essentially collapse the entire economy.  Proportional rationing substantially collapses the economy (with an output below $20\%$ of normal for all countries) and the mixed scheme reduces output for Germany and Spain by more than 70\% and about $60\%$ for Italy. Interestingly, all feasible market allocations lie close to the identity line, suggesting similar impacts to output and consumption.

On a sectoral level, we find the overall orderings of economic impacts are highly correlated across methods and countries, despite a few pronounced differences. An interesting example is Forestry (A02) which represents the only industry that experiences less adverse impacts under proportional rationing compared to mixed rationing. Due to being non-essential and having a low Remote Labor Index (Appendix~\ref{apx:shocks}, Table \ref{t:ind_supply}), forestry is the industry with the largest supply shocks. Thus, it virtually always faces stronger supply constraints than demand constraints. In case of no prioritization of industries, Forestry receives less demand and thus creates smaller input bottlenecks for downstream industries. This, in turn, leads to smaller input bottlenecks for Forestry as its downstream industries can be found upstream as well.

There is also substantial variation in how close industries are to their theoretical maximum value as determined by the optimization method. When considering the proportional rationing algorithm in Italy, for example, the output levels of the best faring industries (such as Education (P85) or Telecommunications (J61)) reach around 30\% of their theoretical maximum. On the other hand, the hardest hit industry, Accommodation-Food (I), is only at a 5\% level of what is possible. Thus, Accommodation-Food is not only hit extremely hard by direct shocks, but also experiences large higher-order effects through the rationing dynamics.

Fig. \ref{fig:res_aggregate} makes it clear that the behavioral assumptions imposed on suppliers matter enormously for economic impact predictions. 
In fact, results vary more across alternative assumptions than across countries.

In the absence of further modeling refinements (such as inventories, adaptive behavior, substitution effects), shock amplification is always pronounced if basic national accounting identities are required to hold. But the actual extent of shock amplification depends strongly on how input bottlenecks are created and passed on downstream in the supply chain.

\subsubsection{Shock magnitude effects} \label{sec:magnitude}

We next investigate the sensitivity of economic impact predictions with respect to shock magnitude and network connectedness. While we have seen in Section \ref{sec:results} that different assumptions on rationing behavior can lead to entirely different estimates of impact, it is not clear whether these results are specific to the three datasets considered. We therefore conduct a series of simulation experiments to gain a better understanding of the generality of the results.

First, we investigate how the estimates of the various methods depend on the magnitude of shocks. To do this, we rescale supply and demand shocks and we apply the optimization methods and the rationing algorithms to the new shock data. We then redo this analysis for various shock scales. To better differentiate between the qualitative effects of demand and supply shock propagation, we allow for different scaling factors for demand and supply constraints, i.e.
\begin{align}
x_{i}^\text{max} &= ( 1-  \alpha^S \epsilon_i^S ) x_{i,0} ,  \\
f_{i}^\text{max} &= ( 1-  \alpha^D \epsilon_i^D ) f_{i,0},
\end{align}
where $\alpha^S, \alpha^D \in [0,1]$.

\begin{figure}[H]
	\centering
	\includegraphics[width=1\textwidth]{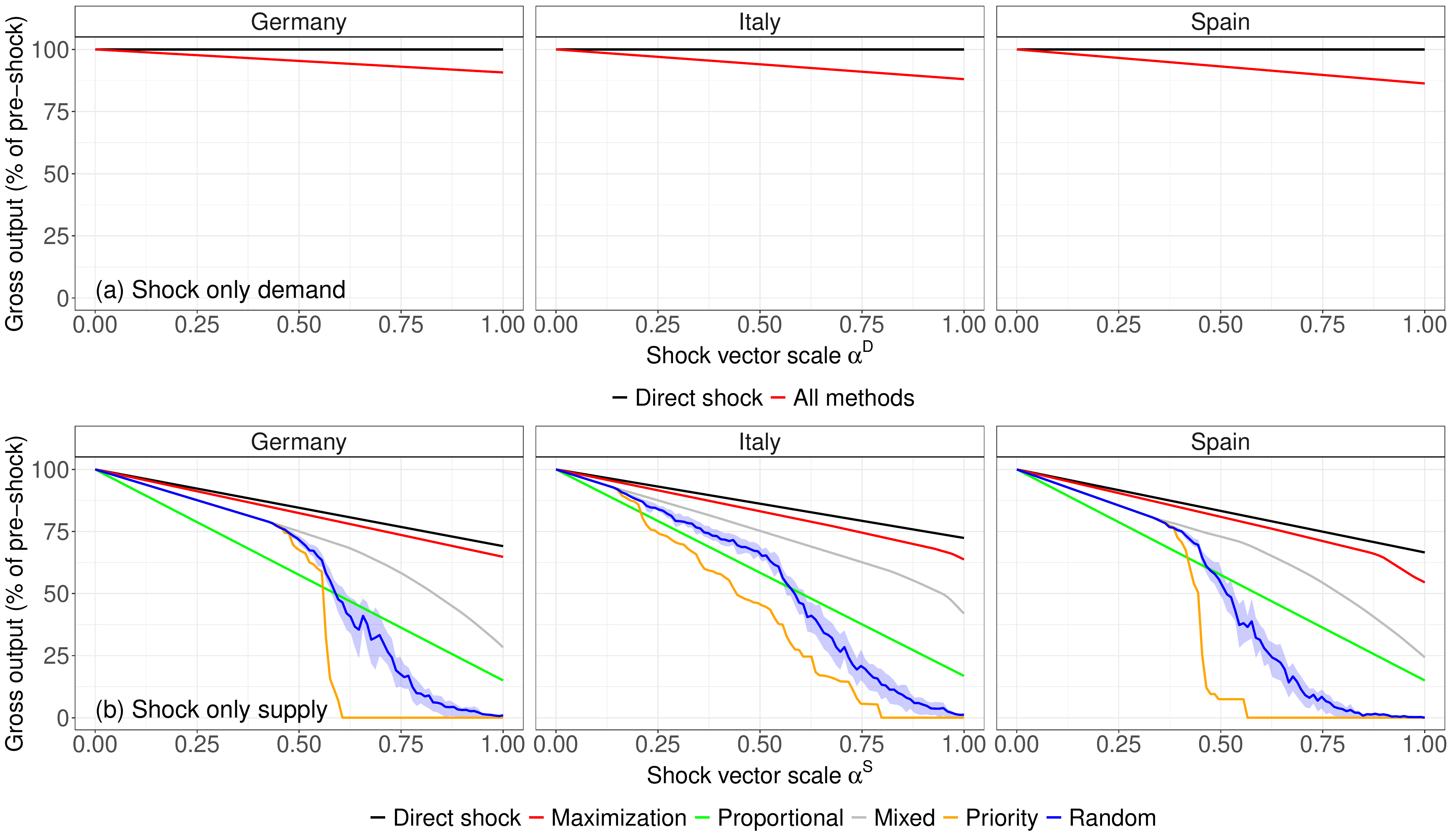}
	\caption{
		{\bf Economic impact as a function of shock magnitude.}
		(a) Aggregate gross output levels as a function of scaling only demand shocks between zero and one ($\alpha^S = 0, \alpha^D \in [0,1]$). All methods yield the same economic impact estimates which scale linearly with $\alpha^S$. The black horizontal line indicates that there is no adverse economic impact from supply side constraints.
		(b) Aggregate gross output levels as a function of scaling only supply shocks between zero and one ($\alpha^D = 0, \alpha^S \in [0,1]$). 
		The \emph{Random rationing} line represents the average taken from 100 samples and the shades indicate the interquartile range.
		We do not distinguish further between the two maximization methods as results are almost identical.
	}
	\label{fig:res_scale_out}
\end{figure}

Fig. \ref{fig:res_scale_out}(a) shows aggregate output levels for all cases when only demand shocks are scaled between zero and one and when there are no further supply constraints being present ($\alpha^S = 0, \alpha^D \in [0,1]$).\footnote{
Since results for aggregate final consumption and aggregate gross output are very similar, we only present figures of the latter in this section. We show results for scaling supply and demand shocks concurrently ($\alpha^S = \alpha^D \in [0,1]$) in Appendix \ref{apx:magnitude}.
} 
It becomes evident that predictions made by the rationing algorithms and the optimization methods are identical and scale linearly with the demand shock magnitude. Thus, the rationing algorithms do not differ with respect to upstream shock propagation and always arrive at the optimal solution in absence of further supply side constrictions.

The picture becomes very different when rescaling only supply shocks and turning off demand shocks ($\alpha^D = 0, \alpha^S \in [0,1]$) as demonstrated in Fig. \ref{fig:res_scale_out}(b). Here, the direct impact reduces gross output linearly as the shock magnitude is increased and puts an upper bound to the solutions of the methods considered here. When there are no shocks, $\alpha^S=0$, all methods recover the empirical IO data as expected. For small shocks we observe that estimated impacts are very similar across different methods, but the proportional rationing algorithm consistently returns the smallest output values.  The results of the other algorithms (mixed, priority, random) lie between those proportional rationing and optimization predictions, and are identical for small shocks up to about $\alpha^S = 0.5$.  As $\alpha^S$ is increased further they deviate dramatically.  Under the priority algorithm even a small increase in $\alpha^S$ causes the economy to collapse.  The random algorithm also causes the economy to collapse, though more slowly, and the mixed algorithm is substantially better.

Fig. \ref{fig:res_scale_out} thus makes clear that the ranking of the impact assessment methods as seen in Fig. \ref{fig:res_aggregate} is not generic, but strongly depends on the shock magnitude. If there are only small supply shocks present, a priority rationing rule is better than proportional rationing. In this case most of the demand of downstream industries can be met and small shocks only propagate to few customers. 
In a proportional rationing setup, on the other hand, shocks are passed on to every customer. Despite small shock magnitudes, the wider breadth of shock propagation leads to comparatively larger aggregate impacts. 

In contrast, priority rationing exacerbates shock propagation compared to proportional rationing in case of large supply shocks.
If industries are severely output constrained and ration based on a priority rule, the demand of several customers' might not be satisfied at all. The Leontief production function, in turn, implies a complete shutdown of these downstream industries due to the input bottlenecks created. More input bottlenecks will be created in further rounds of shock propagation, potentially causing massive collapses. If supply shocks are that large, shock amplification is milder if passed on proportionally to customers.
While here every customer experiences some shocks from a constrained supplier, this effect is outweighed by the fact that proportional rationing avoids the creation of even larger idiosyncratic input bottlenecks.

\subsubsection{Network density} \label{sec:nw_effects}

Intuitively, the extent of shock propagation not only depends on the size of initial shocks but also on how industries are connected with each other. If the production network is dense, i.e. most of the potential links are present, idiosyncratic shocks will spread out very quickly to many other industries in the network. In contrast, if the network is sparse, shock propagation might be more local, at least initially, and takes more steps to spread out in the network.

We therefore conduct an experiment where we again apply the different shock propagation mechanisms to the data, but control for the IO network density. We do this in the following way. First, we randomly eliminate a given number of links in the intermediate consumption matrix $\matr Z$.\footnote{
We also experimented with eliminating smallest links first instead of randomly selection. Results are similar to the ones presented here and shown in the Appendix \ref{apx:nw_effects}.
}
We only consider deleting links instead of adding links since the aggregate IO networks we are using are highly dense ($\sum_{ij} \mathds{1}_{\{z_{ij}>0\} }/ n^2 > 99\%$).
Note that setting a link $z_{ij}>0$ equal to zero without changing final consumption values reduces total output of supplier $i$, since the accounting identity $x_i = \sum_j z_{ij} + f_i$ has to hold. The output of customer $j$ will not be affected if we assume that the reduction in intermediate consumption is absorbed by a respective increase of $j$'s value added (which does not affect the simulations). 
After deleting nodes we thus rebalance the economic system by adjusting gross output of the relevant industries such that the basic national accounting identity is satisfied.
Next, we recompute the technical coefficient matrix $\matr A$, the Leontief matrix $\matr L$ and the vector of productive constraints $\bm x^\text{max}$ to initialize the optimization and rationing simulations. We repeat this procedure 50 times for a given level of density and explore the whole range of possible density values.

\begin{figure}[H]
	\centering
	\includegraphics[width=1\textwidth]{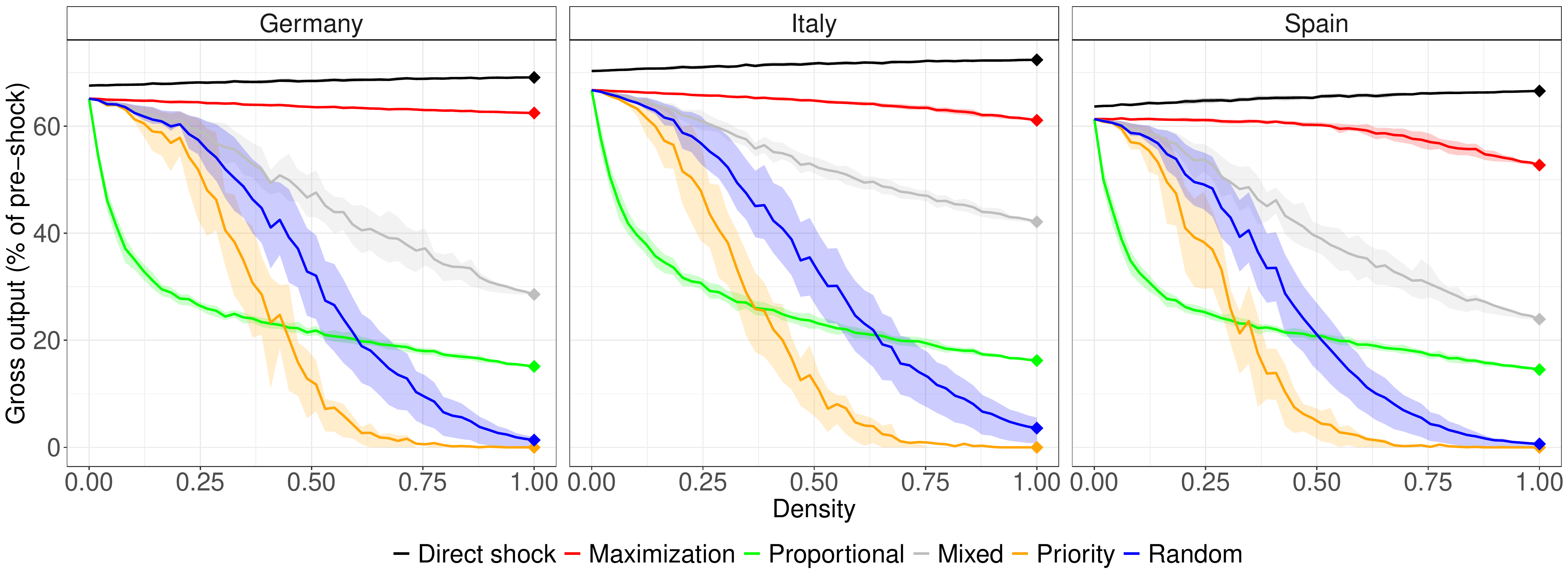}
	\caption{
		{\bf Economic impact as a function of network density.}
		The network density is changed by randomly eliminating links in the IO network. Solid lines are predicted output values obtained from averaging over 50 samples and shades indicate the interquartile range. Since random rationing is stochastic by itself, we apply this algorithm 50 times to every network sample and take averages and quantiles over the full (pooled) density-specific sample.
		The actual data corresponds to the very right of the x-axis, as indicated by the diamonds.
	}
	\label{fig:res_density_rand_out}
\end{figure}

We visualize aggregate output levels predicted by the various impact assessment methods as a function of network density in Fig. \ref{fig:res_density_rand_out}. The plot again confirms that the observed ranking of methods in Fig. \ref{fig:res_aggregate} is not generic but is strongly affected by the network density. It becomes clear that shock amplification is comparatively small if the network is very sparse. On the other hand, if the network is very dense, as the empirical data would suggest, economic impacts are substantial for all cases. In particular, the gap between the optimal solution and the bottom-up rationing approaches widens with higher network density. 
Interestingly, the minimal amplification of direct shocks also increases with higher density values, although the size of this effect is rather small.

If industries have only few suppliers and customers, proportional rationing yields by far the most pessimistic predictions of aggregate impact. If there are many connections among industries, on the contrary, proportional rationing mitigates shock propagation dynamics compared to priority and random rationing. Shock amplification is always mildest if industries use a mixed proportional/priority rationing rule, although even in this case economic impacts increase substantially with higher density.

These results indicate that estimates of economic impact are highly sensitive with respect to the network structure which, in turn, often depends on the level of data aggregation. More aggregated data necessarily implies higher levels of density. For example, industry-level links are the result of pooling many firm links. Firm-level production networks, on the other hand, are an aggregation of many individual contractual relationships and payments. Imagine applying the rationing mechanisms to a firm-level production network in two ways: first to the actual firm-level data and second to an industry aggregate of the same data. Our results suggest that predicted economic impacts could be substantially larger in the second case, despite using the same underlying data.

Our findings also point out that economic impact predictions can be sensitive with respect to data quality. Many disaggregate firm-level production network datasets are substantially biased, as they frequently include only specific supplier-customer relationships which are subject to specific reporting rules. Thus, we would expect real world production networks to be substantially denser than what the data suggests. Our analysis indicates that such biases could have important consequences for impact assessment.

\section{Discussion} \label{sec:discuss}

We have shown that existing IO models have difficulties in dealing with simultaneous supply and demand shocks. 
However, both types of shocks play an important role in situations such as natural disasters and during a pandemic. 
We have introduced a simple optimization method that allows us to find best case market allocations that are consistent with the exogenous shocks. To obtain more realistic, bottom-up impact estimates, we studied alternative rationing dynamics which differ in how suppliers serve customers in case of output constraints. Using IO data for Germany, Italy and Spain, we found that these bottom-up approaches lead to substantial amplifications of initial shocks, which are much higher than optimal solutions would suggest. We further established that different rationing assumptions can lead to dramatically different economic impact predictions. Moreover, these predictions are highly sensitive with respect to the magnitude of first-order shocks and the production network structure.

It is clear that adequate macroeconomic predictions of pandemic impacts require more sophisticated modeling techniques than the ones studied here. Yet the downside of more complicated models is that the underlying mechanisms of predictions can be difficult to isolate. 
We therefore studied relatively simple economic models to better carve out key mechanisms of shock propagation in twofold constrained production networks. 
The choice for simplicity in this study also entails limitations which are important to bear in mind. We did not depart from the assumption of industry-specific Leontief production functions throughout our analysis, although the choice of production function has been shown to be a key variable for impact assessment \citep{pichler2020production}. While input substitutions might be limited in the short-run, fixed production recipes are nevertheless a strong assumption, in particular for the aggregate data considered here.

It is also important to stress that we did not take any adaptive behavior of economic agents into account and did not allow for the possibility of inventory buildup and depletion. By imposing feasible market solutions, we essentially forced the economy to converge into a new equilibrium. It is not clear, however, that a perturbed complex system such as national economies would quickly approach a new equilibrium state instead of following transient paths for an extended period. In general, we refrained from making the time dimension explicit, although we acknowledge its relevance for shock propagation dynamics.

Despite the caveats mentioned, our analysis makes clear that level of aggregation and data quality play an important role in aggregate predictions of shock amplification. Detailed and high-quality production network data are rare, necessitating researchers to study biased or aggregate data. Inevitably, this will affect the connectedness of economic agents and thus influence shock propagation mechanisms. For example, studies based on large-scale firm level databases report network density values well below 0.01\% \citep{kumar2021distress, borsos2020unfolding, peydro2020production, demir2018financial, huneeus2018production, spray2017reorganise}.
We should also mention that estimates of direct shocks to supply and demand are subject to large uncertainties. This could have important consequences for model predictions, due to the sensitivity of shock propagation mechanisms with respect to first-order shock magnitudes.

We find that the number of links between industries strongly influences how different behavior rules amplify direct shocks. The level of connectedness is a very simple aggregate network measure, and we would expect that further aspects, such as community structure, degree/strength heterogeneity or (dis-) assortative mixing, play an important role too. Understanding how alternative rationing assumptions interact with structural properties of complex networks would require more detailed production network data, e.g. at the firm level, and could be an interesting avenue for future research.

We conclude by stressing that our analysis is not constrained to pandemic shocks. While the Covid-19 pandemic is a main motivation of our study, simultaneous supply and demand constraints are ubiquitous features of any economy, in particular in the short run.
Supply shocks are a prominent characteristic of many natural hazards (floods, earthquakes, hurricanes) and tools of mostly demand-driven IO analysis are frequently applied for impact assessment. Our results indicate that adequately modeling shock propagation in production networks will require a better integration of both types of economic constraints.

\FloatBarrier


\FloatBarrier
\normalsize

\newpage
\appendix

\section*{Appendix}

\section{Details on first-order shocks to supply and demand}
\label{apx:shocks}

\begin{table}[ht]
	\scriptsize
	\centering
	\begin{tabular}{|rl|rrr|rrr|}
		\hline
		& & \multicolumn{3}{c|}{$f_i$ (\%)} & \multicolumn{3}{c|}{$\epsilon_i^D$ (\%)}   \\
		ISIC & Industry & DEU & ESP  & ITA & DEU & ESP  & ITA  \\ 
		\hline
		A01 & Agriculture & 0.5 & 1.4 & 1.0 & 10.0 & 9.9 & 10.0 \\ 
		A02 & Forestry & 0.1 & 0.0 & 0.1 & 10.0 & 8.6 & 3.8 \\ 
		A03 & Fishing & 0.0 & 0.1 & 0.0 & 10.0 & 10.0 & 10.0 \\ 
		B & Mining & 0.3 & 0.5 & 0.4 & 10.0 & 10.0 & 10.0 \\ 
		C10-12 & Manuf. Food-Beverages & 4.1 & 5.0 & 4.0 & 10.0 & 10.0 & 10.0 \\ 
		C13-15 & Manuf. Textiles & 0.6 & 1.4 & 2.9 & 10.0 & 10.0 & 10.0 \\ 
		C16 & Manuf. Wood & 0.4 & 0.1 & 0.3 & 10.0 & 10.0 & 9.9 \\ 
		C17 & Manuf. Paper & 0.6 & 0.4 & 0.5 & 10.0 & 10.0 & 10.0 \\ 
		C18 & Media print & 0.1 & 0.1 & 0.1 & 10.0 & 9.9 & 10.0 \\ 
		C19 & Manuf. Coke-Petroleum & 1.7 & 2.6 & 1.4 & 10.0 & 10.0 & 10.0 \\ 
		C20 & Manuf. Chemical & 3.4 & 2.1 & 1.5 & 9.9 & 9.9 & 10.0 \\ 
		C21 & Manuf. Pharmaceutical & 1.1 & 1.0 & 1.2 & 8.1 & 9.3 & 9.8 \\ 
		C22 & Manuf. Rubber-Plastics & 1.4 & 0.7 & 1.0 & 10.0 & 9.9 & 10.0 \\ 
		C23 & Manuf. Minerals & 0.6 & 0.5 & 0.7 & 10.0 & 10.0 & 10.0 \\ 
		C24 & Manuf. Metals-basic & 1.5 & 1.3 & 1.4 & 10.0 & 10.0 & 10.0 \\ 
		C25 & Manuf. Metals-fabricated & 1.8 & 1.0 & 1.6 & 10.0 & 10.0 & 10.0 \\ 
		C26 & Manuf. Electronic & 2.0 & 0.4 & 0.9 & 9.9 & 9.9 & 10.0 \\ 
		C27 & Manuf. Electric & 2.3 & 0.8 & 1.4 & 10.0 & 10.0 & 10.0 \\ 
		C28 & Manuf. Machinery & 5.8 & 1.4 & 4.7 & 10.0 & 10.0 & 10.0 \\ 
		C29 & Manuf. Vehicles & 8.2 & 3.7 & 2.1 & 10.0 & 10.0 & 10.0 \\ 
		C30 & Manuf. Transport-other & 1.0 & 1.1 & 1.0 & 9.9 & 9.9 & 10.0 \\ 
		C31-32 & Manuf. Furniture & 1.4 & 0.6 & 1.4 & 9.9 & 9.9 & 10.0 \\ 
		C33 & Repair-Installation & 0.5 & 0.3 & 0.6 & 10.0 & 10.0 & 10.0 \\ 
		D35 & Electricity-Gas & 1.6 & 1.6 & 1.1 & 2.3 & 0.8 & 1.2 \\ 
		E36 & Water & 0.2 & 0.4 & 0.3 & 1.5 & 0.9 & 0.6 \\ 
		E37-39 & Sewage & 0.7 & 0.7 & 0.6 & 3.6 & 2.5 & 1.7 \\ 
		F & Construction & 5.5 & 7.6 & 7.5 & 10.0 & 9.9 & 9.9 \\ 
		G45 & Vehicle trade & 0.9 & 1.7 & 1.4 & 10.0 & 10.0 & 10.0 \\ 
		G46 & Wholesale & 3.4 & 4.1 & 4.6 & 9.8 & 9.5 & 9.9 \\ 
		G47 & Retail & 3.4 & 5.5 & 5.7 & 9.5 & 9.5 & 9.8 \\ 
		H49 & Land transport & 0.8 & 1.7 & 1.8 & 56.9 & 38.8 & 55.2 \\ 
		H50 & Water transport & 0.7 & 0.2 & 0.5 & 11.4 & 27.5 & 47.3 \\ 
		H51 & Air transport & 0.6 & 0.6 & 0.4 & 50.3 & 24.7 & 47.3 \\ 
		H52 & Warehousing & 0.3 & 0.8 & 1.1 & 22.3 & 19.7 & 33.5 \\ 
		H53 & Postal & 0.1 & 0.0 & 0.1 & 3.4 & 2.2 & 3.4 \\ 
		I & Accommodation-Food & 2.4 & 8.6 & 4.7 & 73.1 & 75.5 & 79.1 \\ 
		J58 & Publishing & 0.4 & 0.3 & 0.3 & 3.8 & 5.4 & 4.0 \\ 
		J59-60 & Video-Sound-Broadcasting & 0.6 & 0.5 & 0.3 & 4.3 & 3.8 & 3.5 \\ 
		J61 & Telecommunications & 0.9 & 1.3 & 1.2 & 1.1 & 1.6 & 3.0 \\ 
		J62-63 & IT & 1.5 & 1.6 & 1.1 & 8.8 & 9.5 & 8.5 \\ 
		K64 & Finance & 1.9 & 0.8 & 0.9 & 2.9 & 3.8 & 1.6 \\ 
		K65 & Insurance & 1.4 & 1.1 & 1.0 & 1.3 & 0.8 & 1.1 \\ 
		K66 & Auxil. Finance-Insurance & 0.0 & 0.2 & 0.1 & 2.0 & 1.9 & 4.9 \\ 
		L68 & Real estate & 7.2 & 7.8 & 9.8 & 0.2 & 0.1 & 0.5 \\ 
		M69-70 & Legal & 0.7 & 0.6 & 0.4 & 9.3 & 7.9 & 5.2 \\ 
		M71 & Architecture-Engineering & 1.0 & 1.0 & 0.2 & 9.4 & 9.3 & 8.4 \\ 
		M72 & R\&D & 0.9 & 0.5 & 0.6 & 8.3 & 7.9 & 9.8 \\ 
		M73 & Advertising & 0.1 & 0.1 & 0.1 & 10.0 & 9.1 & 9.2 \\ 
		M74-75 & Other Science & 0.3 & 0.1 & 0.3 & 3.5 & 3.5 & 5.0 \\ 
		N & Private Administration & 1.2 & 1.2 & 1.1 & 4.2 & 4.2 & 4.9 \\ 
		O84 & Public Administration & 6.1 & 6.3 & 7.1 & 0.2 & 0.7 & 0.0 \\ 
		P85 & Education & 4.1 & 5.1 & 3.9 & 0.7 & 1.0 & 0.0 \\ 
		Q & Health & 8.3 & 7.2 & 7.5 & 0.1 & 0.1 & 0.1 \\ 
		R\_S & Other Service & 3.2 & 3.3 & 3.2 & 4.3 & 4.3 & 4.7 \\ 
		T & Household activities & 0.2 & 0.8 & 1.1 & 0.0 & -0.0 & -0.0 \\
		\hline
	\end{tabular}
	\caption{ \textbf{Industry-specific demand shock details.} $f_i$ denotes final consumption per industry as fraction of aggregate final consumption. $\epsilon_i^D$ is the total demand shock per industry.  }
	\label{t:ind_demand}
\end{table}

\begin{table}[ht]
	\scriptsize
	\centering
	\begin{tabular}{|rl|rrr|rrr|rrr|r|}
		\hline
		& & \multicolumn{3}{c|}{$x_i$ (\%)} & \multicolumn{3}{c|}{$\epsilon_i^S$ (\%)} & \multicolumn{3}{c|}{$e_i$ (\%)} &  $\text{RLI}_i$\\
		ISIC & Industry & DEU & ESP  & ITA & DEU & ESP  & ITA & DEU & ESP  & ITA &  (\%) \\ 
		\hline
		A01 & Agriculture & 0.9 & 2.2 & 1.7 & 0.0 & 0.0 & 0.0 & 100.0 & 100.0 & 100.0 & 13.6 \\ 
		A02 & Forestry & 0.1 & 0.0 & 0.0 & 85.0 & 85.0 & 85.0 & 0.0 & 0.0 & 0.0 & 15.0 \\ 
		A03 & Fishing & 0.0 & 0.1 & 0.1 & 0.0 & 0.0 & 0.0 & 100.0 & 100.0 & 100.0 & 35.7 \\ 
		B & Mining & 0.2 & 0.3 & 0.3 & 48.3 & 48.3 & 34.5 & 30.0 & 30.0 & 50.0 & 31.0 \\ 
		C10-12 & Manuf. Food-Beverages & 3.5 & 6.9 & 4.1 & 0.0 & 26.0 & 26.0 & 100.0 & 66.7 & 66.7 & 22.1 \\ 
		C13-15 & Manuf. Textiles & 0.4 & 1.0 & 2.6 & 68.5 & 68.5 & 59.4 & 0.0 & 0.0 & 13.3 & 31.5 \\ 
		C16 & Manuf. Wood & 0.5 & 0.3 & 0.5 & 73.1 & 73.1 & 60.6 & 0.0 & 0.0 & 17.0 & 26.9 \\ 
		C17 & Manuf. Paper & 0.7 & 0.6 & 0.7 & 34.3 & 0.0 & 48.7 & 50.0 & 100.0 & 29.0 & 31.5 \\ 
		C18 & Media print & 0.4 & 0.4 & 0.4 & 0.0 & 0.0 & 0.0 & 100.0 & 100.0 & 100.0 & 39.0 \\ 
		C19 & Manuf. Coke-Petroleum & 1.5 & 2.4 & 1.7 & 0.0 & 6.4 & 0.0 & 100.0 & 90.0 & 100.0 & 36.0 \\ 
		C20 & Manuf. Chemical & 2.6 & 2.5 & 1.6 & 19.0 & 52.5 & 8.2 & 70.0 & 17.0 & 87.0 & 36.7 \\ 
		C21 & Manuf. Pharmaceutical & 0.9 & 0.7 & 0.8 & 0.0 & 0.0 & 0.0 & 100.0 & 100.0 & 100.0 & 40.3 \\ 
		C22 & Manuf. Rubber-Plastics & 1.4 & 0.9 & 1.3 & 35.3 & 70.5 & 47.2 & 50.0 & 0.0 & 33.0 & 29.5 \\ 
		C23 & Manuf. Minerals & 0.8 & 0.8 & 1.0 & 63.9 & 63.9 & 61.4 & 0.0 & 0.0 & 4.0 & 36.1 \\ 
		C24 & Manuf. Metals-basic & 1.9 & 2.1 & 1.8 & 72.6 & 72.6 & 72.6 & 0.0 & 0.0 & 0.0 & 27.4 \\ 
		C25 & Manuf. Metals-fabricated & 2.4 & 1.4 & 2.6 & 66.3 & 66.3 & 66.3 & 0.0 & 0.0 & 0.0 & 33.7 \\ 
		C26 & Manuf. Electronic & 1.4 & 0.4 & 0.7 & 43.1 & 43.1 & 38.8 & 0.0 & 0.0 & 10.0 & 56.9 \\ 
		C27 & Manuf. Electric & 1.9 & 0.8 & 1.1 & 63.1 & 63.1 & 50.5 & 0.0 & 0.0 & 20.0 & 36.9 \\ 
		C28 & Manuf. Machinery & 4.5 & 1.2 & 3.6 & 61.8 & 61.8 & 47.0 & 0.0 & 0.0 & 24.0 & 38.2 \\ 
		C29 & Manuf. Vehicles & 6.3 & 2.5 & 1.5 & 69.7 & 69.7 & 69.7 & 0.0 & 0.0 & 0.0 & 30.3 \\ 
		C30 & Manuf. Transport-other & 0.8 & 0.8 & 0.7 & 59.7 & 59.7 & 59.7 & 0.0 & 0.0 & 0.0 & 40.3 \\ 
		C31-32 & Manuf. Furniture & 0.9 & 0.6 & 1.2 & 65.2 & 59.7 & 58.0 & 0.0 & 8.5 & 11.0 & 34.8 \\ 
		C33 & Repair-Installation & 0.7 & 0.5 & 0.6 & 60.6 & 60.6 & 34.0 & 0.0 & 0.0 & 44.0 & 39.4 \\ 
		D35 & Electricity-Gas & 2.4 & 4.7 & 2.7 & 0.0 & 5.8 & 0.0 & 100.0 & 90.0 & 100.0 & 41.6 \\ 
		E36 & Water & 0.2 & 0.5 & 0.3 & 0.0 & 0.0 & 0.0 & 100.0 & 100.0 & 100.0 & 33.5 \\ 
		E37-39 & Sewage & 0.9 & 0.8 & 1.3 & 0.0 & 0.0 & 0.0 & 100.0 & 100.0 & 100.0 & 29.8 \\ 
		F & Construction & 5.2 & 6.4 & 6.7 & 71.6 & 71.6 & 49.6 & 0.0 & 0.0 & 30.7 & 28.4 \\ 
		G45 & Vehicle trade & 1.1 & 1.4 & 1.1 & 18.0 & 27.3 & 13.7 & 67.0 & 50.0 & 75.0 & 45.4 \\ 
		G46 & Wholesale & 3.9 & 4.9 & 5.3 & 0.0 & 30.0 & 33.5 & 100.0 & 40.0 & 33.0 & 50.1 \\ 
		G47 & Retail & 3.0 & 3.9 & 3.9 & 24.7 & 25.7 & 25.2 & 51.0 & 49.0 & 50.0 & 49.7 \\ 
		H49 & Land transport & 1.8 & 2.5 & 3.0 & 0.0 & 0.0 & 0.0 & 100.0 & 100.0 & 100.0 & 31.4 \\ 
		H50 & Water transport & 0.5 & 0.2 & 0.4 & 0.0 & 0.0 & 0.0 & 100.0 & 100.0 & 100.0 & 35.3 \\ 
		H51 & Air transport & 0.5 & 0.5 & 0.4 & 0.0 & 24.2 & 0.0 & 100.0 & 66.0 & 100.0 & 28.8 \\ 
		H52 & Warehousing & 2.3 & 2.1 & 2.1 & 0.0 & 0.0 & 0.0 & 100.0 & 100.0 & 100.0 & 29.6 \\ 
		H53 & Postal & 0.6 & 0.2 & 0.2 & 0.0 & 0.0 & 0.0 & 100.0 & 100.0 & 100.0 & 35.6 \\ 
		I & Accommodation-Food & 1.6 & 5.8 & 3.3 & 64.6 & 64.6 & 56.6 & 0.0 & 0.0 & 12.5 & 35.4 \\ 
		J58 & Publishing & 0.6 & 0.4 & 0.3 & 0.0 & 7.6 & 0.0 & 100.0 & 75.0 & 100.0 & 69.8 \\ 
		J59-60 & Video-Sound-Broadcasting & 0.6 & 0.6 & 0.5 & 0.0 & 11.0 & 0.0 & 100.0 & 75.0 & 100.0 & 56.1 \\ 
		J61 & Telecommunications & 1.2 & 1.8 & 1.3 & 0.0 & 0.0 & 0.0 & 100.0 & 100.0 & 100.0 & 55.1 \\ 
		J62-63 & IT & 2.1 & 1.4 & 1.6 & 7.2 & 14.4 & 0.0 & 75.0 & 50.0 & 100.0 & 71.1 \\ 
		K64 & Finance & 2.7 & 2.1 & 2.9 & 0.0 & 0.0 & 0.0 & 100.0 & 100.0 & 100.0 & 71.4 \\ 
		K65 & Insurance & 1.4 & 1.0 & 0.8 & 0.0 & 0.0 & 0.0 & 100.0 & 100.0 & 100.0 & 71.3 \\ 
		K66 & Auxil. Finance-Insurance & 0.6 & 0.4 & 1.0 & 0.0 & 0.0 & 0.0 & 100.0 & 100.0 & 100.0 & 71.7 \\ 
		L68 & Real estate & 7.2 & 6.7 & 7.5 & 51.3 & 51.3 & 51.3 & 0.0 & 0.0 & 0.0 & 48.7 \\ 
		M69-70 & Legal & 2.5 & 1.5 & 2.3 & 36.3 & 18.1 & 0.0 & 0.0 & 50.0 & 100.0 & 63.7 \\ 
		M71 & Architecture-Engineering & 1.2 & 1.1 & 1.0 & 45.9 & 45.9 & 0.0 & 0.0 & 0.0 & 100.0 & 54.1 \\ 
		M72 & R\&D & 0.6 & 0.3 & 0.4 & 41.1 & 41.1 & 0.0 & 0.0 & 0.0 & 100.0 & 58.9 \\ 
		M73 & Advertising & 0.4 & 0.5 & 0.5 & 39.7 & 39.7 & 39.7 & 0.0 & 0.0 & 0.0 & 60.3 \\ 
		M74-75 & Other Science & 0.4 & 0.3 & 0.7 & 19.6 & 19.6 & 0.0 & 50.0 & 50.0 & 100.0 & 60.8 \\ 
		N & Private Administration & 3.9 & 2.6 & 3.0 & 42.7 & 54.3 & 41.6 & 33.3 & 15.2 & 35.0 & 36.0 \\ 
		O84 & Public Administration & 4.6 & 4.4 & 4.2 & 0.0 & 0.0 & 0.0 & 100.0 & 100.0 & 100.0 & 44.6 \\ 
		P85 & Education & 2.9 & 3.4 & 2.4 & 0.0 & 46.0 & 0.0 & 100.0 & 0.0 & 100.0 & 54.0 \\ 
		Q & Health & 5.4 & 4.8 & 4.9 & 0.0 & 0.0 & 0.0 & 100.0 & 100.0 & 100.0 & 36.0 \\ 
		R\_S & Other Service & 2.8 & 2.7 & 2.7 & 61.2 & 56.0 & 49.2 & 0.0 & 8.6 & 19.6 & 38.8 \\ 
		T & Household activities & 0.1 & 0.5 & 0.6 & 0.0 & 0.0 & 0.0 & 0.0 & 0.0 & 50.0 & 100.0 \\ 
		\hline
	\end{tabular}
	\caption{ \textbf{Industry-specific supply shock details.} $x_i$ denotes gross output per industry as share of aggregate output. $\epsilon_i^S$ is the total supply shock per industry. $e_i$ is the extent to which an industry is considered as essential following government policies. $\text{RLI}_i$ is the Remote Labor Index.  }
	\label{t:ind_supply}
\end{table}

\FloatBarrier
\section{Details on mixed endogenous/exogenous modeling}
\label{apx:meem}

A core motivation to study shock propagation based on alternative, bottom-up approaches was our observation that existing methods such as the mixed endogenous/exogenous model (MEEM) yields infeasible solution when applied to pandemic shocks.
In the MEEM the $n$ industries are divided into two groups. The first group is the set of $n^s$ supply constrained industries and the second group the $n^d$ demand-constrained industries (we have $n^s + n^d = n$).  We can use this setup to partition Eq.~\eqref{eq:leontief} into
\begin{equation} \label{eq:meem_basic}
\begin{pmatrix}
\bm x^s \\
\bm x^d 
\end{pmatrix} = 
\begin{pmatrix}
\matr A^{ss} & \matr A^{sd} \\
\matr A^{ds} & \matr A^{dd} 
\end{pmatrix}
\begin{pmatrix}
\bm x^s \\
\bm x^d 
\end{pmatrix} +
\begin{pmatrix}
\bm f^s \\
\bm f^d 
\end{pmatrix},
\end{equation}
where superscripts $s$ and $d$ denote supply and demand constrained industries, respectively. The matrix block $\matr A^{sd}$ indicates the input recipes of demand constrained customers with respect to output constrained suppliers and analogously for the other blocks.
In this framework vectors $\bm f^s$ and $\bm x^d$ are endogenous and $\bm f^d$ and $\bm x^s$ exogenous. Rearranging Eq.~\eqref{eq:meem_basic} such that all endogenous variables appear on the left-hand side and all exogenous variables on the right-hand side, yields
\begin{equation} \label{eq:meem_full}
\begin{pmatrix}
\bm f^s \\
\bm x^d 
\end{pmatrix} = 
\begin{pmatrix}
\mathbf{I} & \matr A^{sd} \\
\bm 0 & \mathbf{I} - \matr A^{dd} 
\end{pmatrix}^{-1}
\begin{pmatrix}
\mathbf{I} - \matr A^{ss} & \bm 0 \\
\matr A^{ds} & \mathbf{I} 
\end{pmatrix}
\begin{pmatrix}
\bm x^s \\
\bm f^d 
\end{pmatrix}
\end{equation} 
(for details see \cite{miller2009input}, 621-633).

To apply the MEEM in the pandemic context, we categorize industries into a demand and a supply constrained group based on which shock is larger. If the supply shock of an industry exceeds its demand shock in absolute terms, we treat this industry as supply constrained and vice versa. 
The shock magnitudes for supply and demand for each industry are given as
\begin{align}
x_{i}^\text{SS}  &=  x_{i,0} - x_{i}^\text{max} = - \epsilon_i^S x_{i,0}, \\
f_{i}^\text{DS}  &=  c_{i,0} - f_{i}^\text{max} = - \epsilon_i^D f_{i,0},
\end{align}
where $d_{i}^\text{SS}$ denotes the total \emph{supply shock} and $f_{i}^\text{DS}$ the \emph{demand shock}. If $x_{i}^\text{SS} > f_{i}^\text{DS}$, we consider this industry as supply constrained and we will use its gross output values on the right hand side of Eq.~\eqref{eq:meem_full}. Otherwise we treat it as demand constrained.

\begin{figure}[H]
	\centering
	\includegraphics[width=\textwidth]{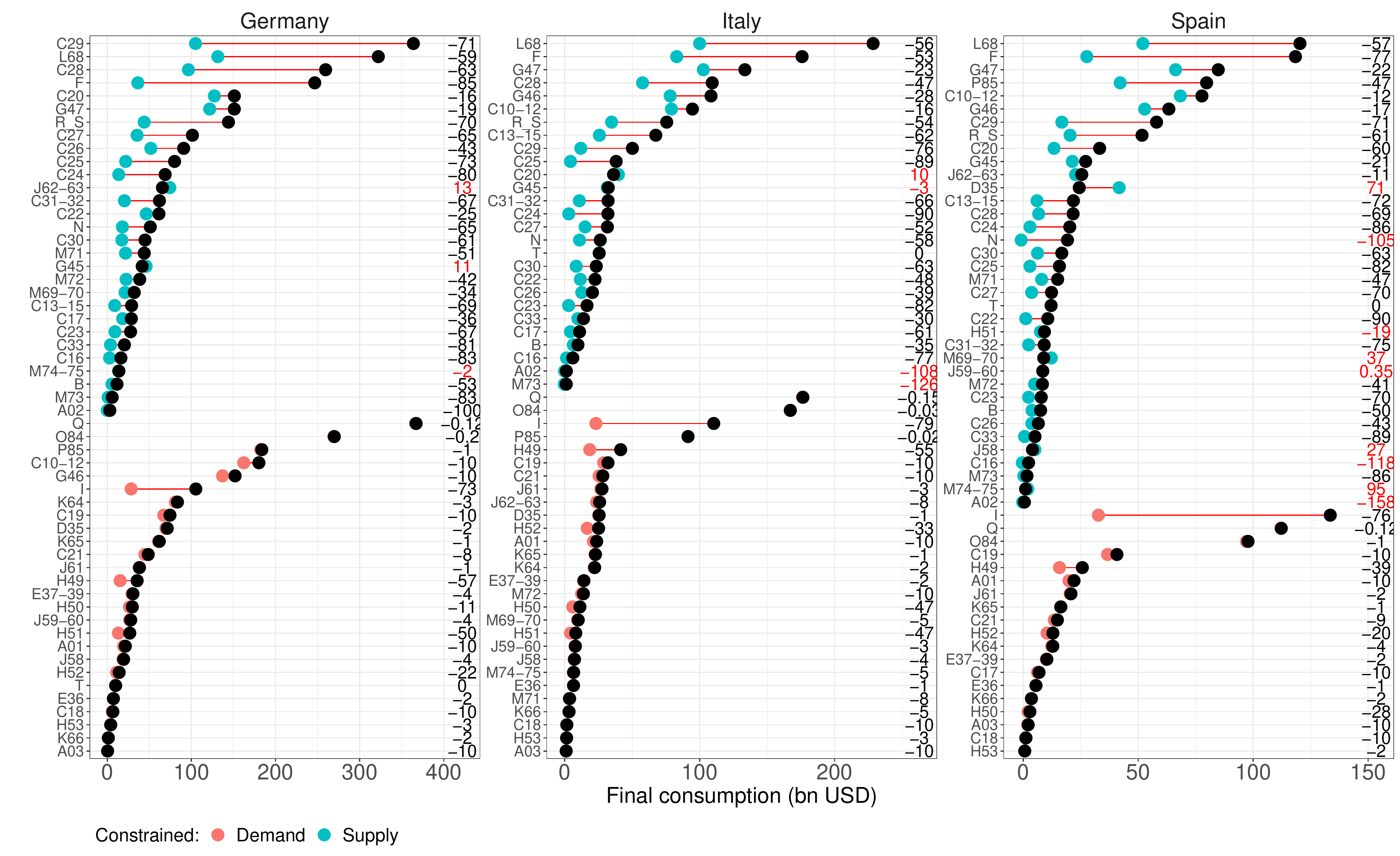}
	\caption{
		{\bf Final consumption values from mixed exogenous/endogenous IO modeling with simultaneous supply and demand shocks.} 
		Black circles indicate initial pre-shock values, red circles demand-constrained industries and blue circles supply-constrained industries. Note that demand-constrained consumption values are exogenously determined by the first-order shocks, while in other cases the change in output depends on the higher order effects on the whole economy.  Red lines indicate the overall change in production for each sector. Numbers to the right of both panels indicate the value change in percentages and are colored red if the result is infeasible.  
	}
	\label{fig:mixedIO_res}
\end{figure}

Following this approach, we apply the MEEM to the IO data of Germany, Italy and Spain by calibrating it to the estimated pandemic supply and demand shocks. 
As shown in Fig. \ref{fig:mixedIO_res}, the MEEM does not yield a feasible solution for any of the three countries. Violations of feasibility conditions are most frequent for Spain, which faces the largest shocks to supply and demand. The model does not compute any negative final consumption values for Germany, but still allocates final consumption values to industries which are larger than $f_i^\text{max}$. Note that the results obtained by the MEEM are out of the solution space delimited by the exogenous constraints on output and consumption. Thus, these results are difficult to interpret in this context and we thus refrain from comparing them more systematically with the results shown in the main text.

The MEEM can yield infeasible economic allocations and it depends on the context whether negative final consumption values are meaningful or not \citep[p.~628]{miller2009input}.
To see why the MEEM can give negative final consumption values, let us consider the output-constrained part of Eq.~\eqref{eq:meem_full}, which can be rewritten as
\begin{align} \label{eq:meem_fs}
\bm f^s = [ \mathbf I - \matr A^{ss} ] \bm x^s - \matr A^{sd} [ \mathbf I - \matr A^{dd}]^{-1} (\matr A^{ds} \bm x^s + \bm f^d).
\end{align}
Note that the vector $(\matr A^{ds} \bm x^s + \bm f^d)$ is always non-negative by definition, as is every element of the matrix $\matr A^{sd} [ \mathbf I - \matr A^{dd}]^{-1}$.  (This is clear from invoking the Hawkins-Simon conditions). Thus, for $\bm f^s \ge \bm 0$, $[ \mathbf I - \matr A^{ss} ] \bm x^s$ must be non-negative and larger than the part to the right of the minus in Eq.~\eqref{eq:meem_fs}. However, this term can be negative for supply shocks that are sufficiently heterogeneous. For example, consider the case of an economy with only two supply-constrained industries without self-loops, $i$ and $j$.  Any supply shocks which lead to $x_i^\text{max} < a_{ij}^{ss} x_{j}^\text{max}$ yield negative final consumption values for industry $i$.
This demonstrates that the MEEM framework is unlikely to yield plausible solutions in the current context.

The MEEM can also lead to final consumption values that lie above any given pre-specified upper limit of consumption $f_i^\text{max}$.  Let us again consider an example of two industries, $i$ and $j$, where $i$ is supply constrained and $j$ is demand constrained. If industry $i$ only supplies industry $j$ and final consumers and industry $j$ only supplies to final consumers, it can be verified that $f_i^s > f_i^\text{max}$ if $x_{i}^\text{SS} - f_{i}^\text{DS} < a_{ij}^{sd} f_j^\text{DS}$. Thus, in case industry $i$ is only slightly supply constrained (supply shocks are only slightly larger than demand shocks) and industry $j$ faces comparatively serious demand constraints, the MEEM would compute larger final consumption values for industry $i$ than are possible.

Note that gross output values of demand constrained industries always lie within the feasible range $x_i^d \in [0,x_i^\text{max}]$ if only adverse supply shocks to the economy are allowed ($\epsilon_i^S \ge 0$). By noting that $\bm x^d =  [ \mathbf I - \matr A^{dd}]^{-1} (\matr A^{ds} \bm x^s + \bm f^d)$, it can easily be verified that $x_i^d \ge 0$. The condition that $\epsilon_i^S x_i^d < \epsilon_i^D f_i^d$ ensures that output of demand constrained industries cannot exceed the maximum output value $x_i^\text{max}$.

The supply and demand shocks do not need to be large for the MEEM to generate infeasible solutions.  
To show this we scale down the size of the supply and demand shocks by varying a parameter $\alpha \in [0,1]$ to obtain new maximum output and consumption values, according to
\begin{align}
x_{i}^\text{max} &= ( 1-  \alpha \epsilon_i^S ) x_{i,0} ,  \\
f_{i}^\text{max} &= ( 1-  \alpha \epsilon_i^D ) f_{i,0}.
\end{align}
Since we scale demand and supply shocks by the same proportion this does not change which industries are supply or demand constrained.

\begin{figure}[H]
	\centering
	\includegraphics[width=1\textwidth]{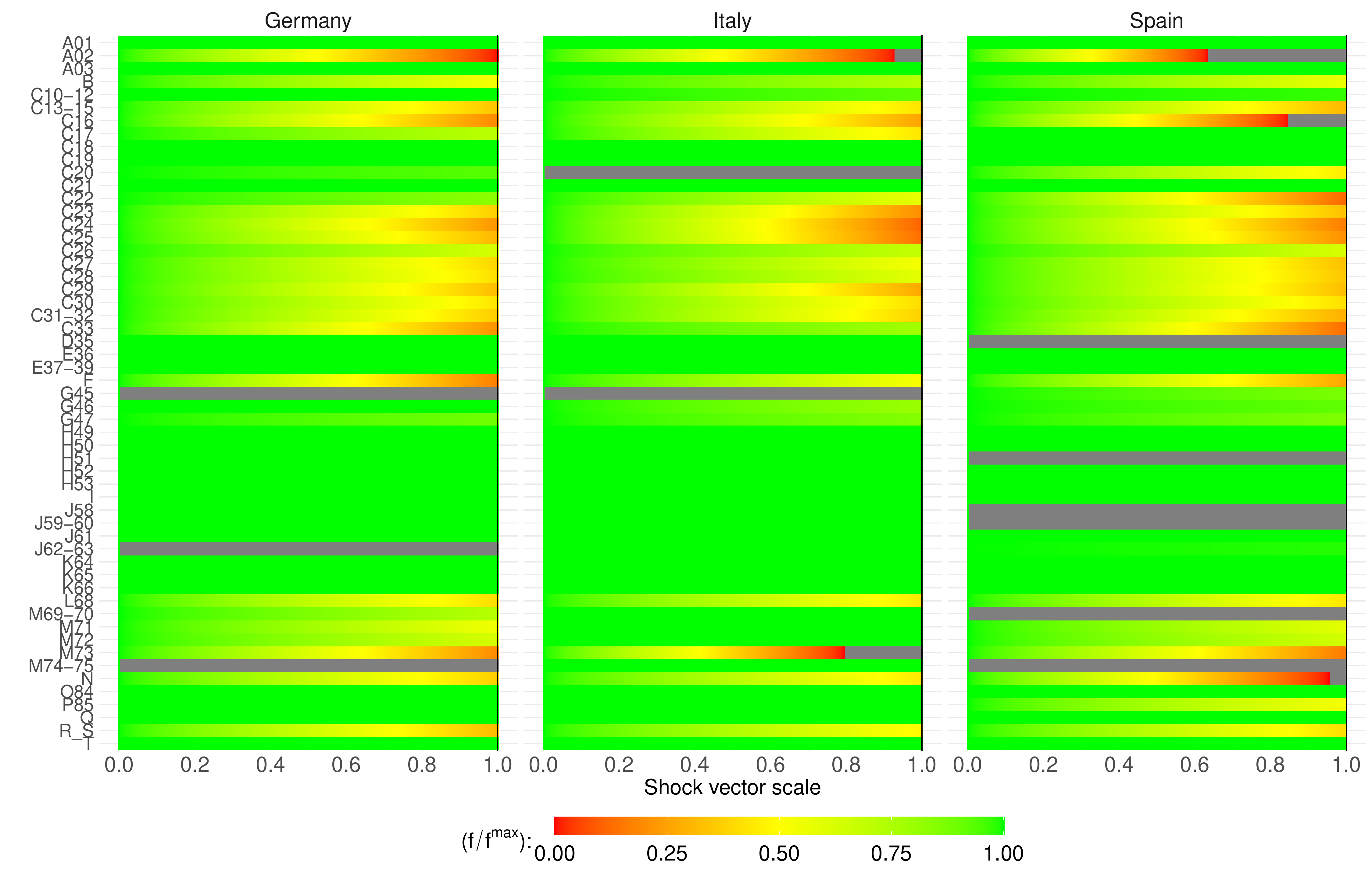}
	\caption{
		{ \bf Infeasible results for final consumption for the MEEM model as a function of shock size.}
		The parameter $\alpha$ that scales the supply and demand shocks varies along the x-axis. The color codes indicate the ratio $f/f^\text{max}$ where $f$ is the MEEM result of final consumption. Note that this ratio is always equal to one (green color) for industries which are demand constrained. Infeasible values, $f \notin [0,f^\text{max}]$, are indicated in gray.
	}
	\label{fig:fshock_scale}
\end{figure}

Fig. \ref{fig:fshock_scale} shows the MEEM results for varying directs shocks. If $\alpha=0$, there is no direct shock, resulting in a feasible market allocation since in this case the MEEM simply recovers the pre-shock economy. This is indicated by the green colors at the very left of all three panels. But even for very small $\alpha>0$ we obtain infeasible solutions for all three countries as shown by the gray colors. If we increase shock sizes further, it becomes more likely that the model computes negative final consumptions values as can be seen from the transition of red colors into gray when following the x-axis from left to right.

\FloatBarrier
\section{Details on shock magnitude effects}
\label{apx:magnitude}

\begin{figure}[H]
	\centering
	\includegraphics[width=1\textwidth]{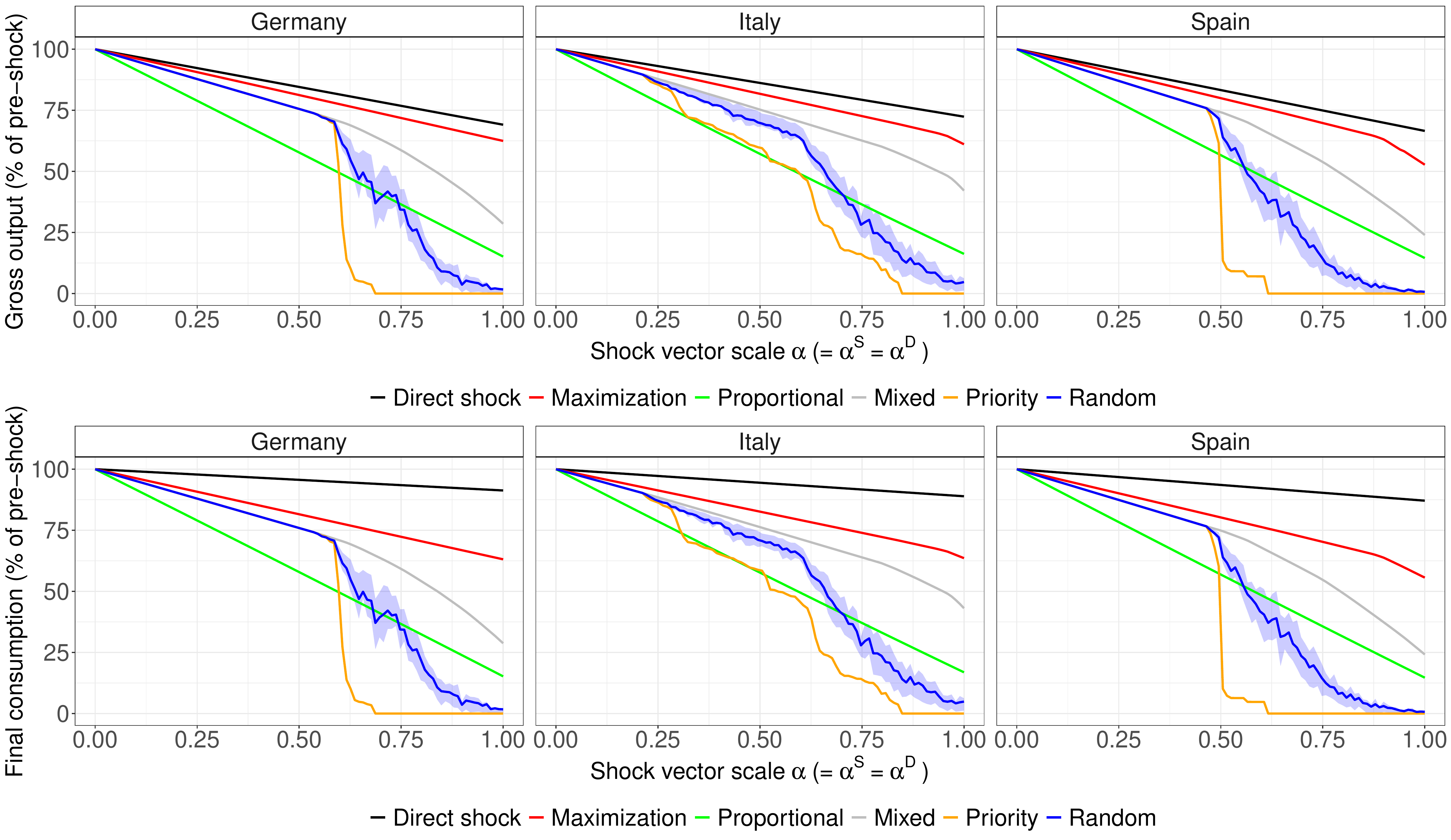}
	\caption{
		{\bf Economic impact as a function of shock magnitude.}
		Aggregate gross output and final consumption levels as a function of scaling demand and supply shocks equally between zero and one ($\alpha = \alpha^S = \alpha^D \in [0,1]$).
	}
	\label{fig:scale_ratio_both}
\end{figure}

In the main text we have shown the effect of scaling either supply or demand shocks on aggregate output. We also explored shock amplification effects when scaling both, supply and demand shocks, simultaneously. Fig. \ref{fig:scale_ratio_both} shows the effect of different shock magnitudes on aggregate output and final consumption values. Results are fairly similar for aggregate values of output and consumption and are also in qualitative agreement with the results presented in Fig. \ref{fig:res_scale_out}.

\FloatBarrier
\section{Details on network density}
\label{apx:nw_effects}

\begin{figure}[H]
	\centering
	\includegraphics[width=1\textwidth]{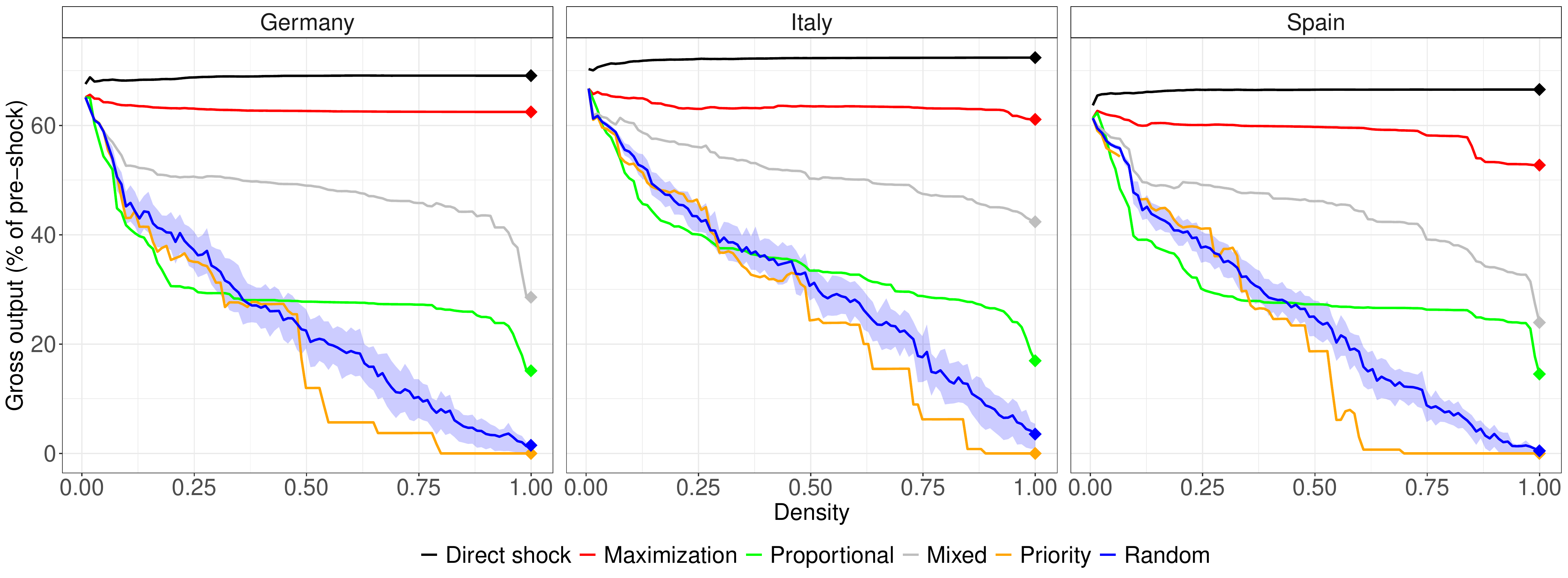}
	\caption{
		{\bf Economic impact as a function of network density.} The figure is the same as Fig. \ref{fig:res_density_rand_out} in the main text, except that network density is changed by eliminating links based on their size instead of random deletion. 
	}
	\label{fig:res_density_small}
\end{figure}

In Section \ref{sec:nw_effects} we removed links randomly and repeated this procedure multiple times for any desired network density value. While random edge removal is one possible approach, other procedures could be followed too. A natural alternative to random edge deletion is to first delete small links. While the high aggregation of the data results in almost complete graphs, link sizes are highly heterogeneous, implying the existence of many very small links \citep{mcnerney2013network, cerina2015world}. It could be argued that many of these links are rather an artifact of data aggregation instead of encoding a fixed production recipe. We therefore repeat the procedure of Section \ref{sec:nw_effects} but eliminate smaller before larger links to achieve a given level of network density. Doing this results in Fig. \ref{fig:res_density_small} which indicates qualitatively similar results as Fig. \ref{fig:res_density_rand_out} of the main text.

\begin{figure}[H]
	\centering
	\includegraphics[width=1\textwidth]{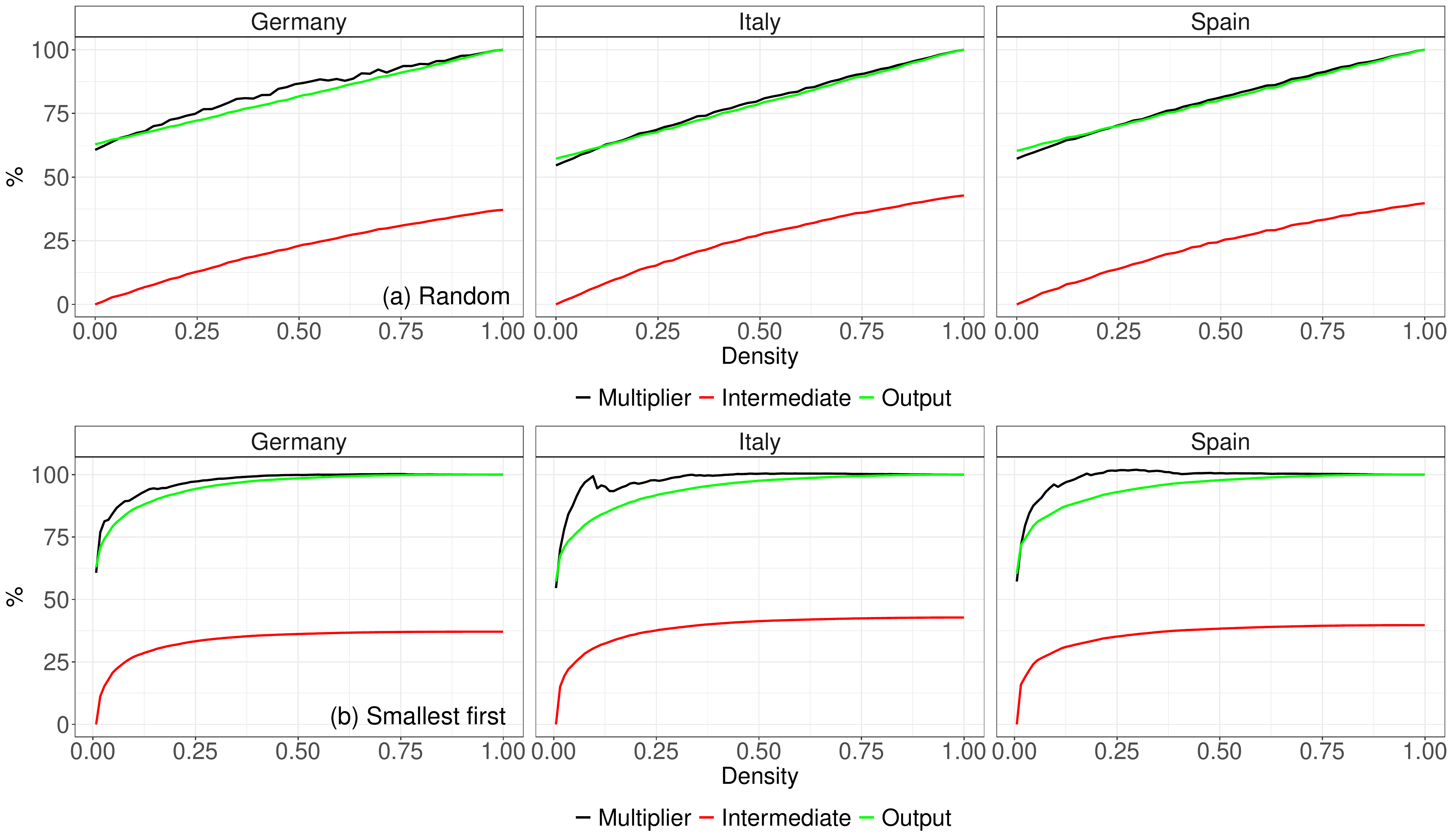}
	\caption{
		{\bf Key economic measures as a function of network density.} 
		(a) The network density is changed by eliminating links randomly (corresponding to Fig. \ref{fig:res_density_rand_out}). 
		(b) The network density is changed by eliminating smaller before larger links (corresponding to Fig. \ref{fig:res_density_small}).
		\emph{Multiplier} refers to the (unweighted) average multiplier of the economy, $\sum_{ij} l_{ij}/n$, after rebalancing as percentage of the initial economy. 
		\emph{Intermediate} denotes the share of intermediate consumption in total output, $\sum_{ij} z_{ij}/ \sum_i x_i $ after rebalancing the economy.
		\emph{Output} is total output after rebalancing divided by initial total output.
	}
	\label{fig:res_density_meta}
\end{figure}

The elimination of existing IO links changes key properties of the underlying economic system. As discussed in Section \ref{sec:nw_effects}, removing intermediate consumption values will reduce aggregate output. Figs. \ref{fig:res_density_meta}(a) and \ref{fig:res_density_meta}(b) visualize the relationship between output and network density following the random link and ``smallest-first'' removal approach, respectively.
Similarly, the ratio intermediate consumption over aggregate output will be reduced as a consequence. A density value equal to zero means that there is no intermediate consumption left and firms only use primary factors as inputs in production. Fig. \ref{fig:res_density_meta} also shows that the average output multiplier decreases when making the network sparser, although not necessarily monotonously.
Overall, the economic indicators change fairly linearly with respect to network density if links are randomly removed, whereas these relationships are highly nonlinear if smaller edges are eliminated before larger ones.

\end{document}